\documentclass[12pt]{article}
\usepackage{amsmath,epsfig,a4,cite,latexsym,axodraw}
\usepackage{mathrsfs}
\usepackage{feynmp}               
\def\pslash#1{{\setbox0=\hbox{$#1$}
  \rlap{\ifdim\wd0>.7em\kern.22\wd0\else\kern.1\wd0\fi /}#1}}
\def\psl{\pslash p}

\def\Dsl{\pslash D}
\def\partialsl{\pslash \partial}

\def\MSbar{{$\overline{\mbox{MS}}$}}

\def\thetaDRED{\theta_{\rm DRED}}
\def\AA{{\cal A}}
\def\BB{{\cal B}}

\def\tlgenuine{{\rm 2L\,diag\,(\gamma)}}
\def\chipmphot{{\chi^\pm\,(\gamma)}}
\def\chizphot{{\chi^0\,(\gamma)}}
\def\SUSYtlphot{{\rm SUSY,2L\,(\gamma)}}
\def\ctQED{{\rm ct,QED}}
\def\ctSUSY{{\rm ct,SUSY}}

\def\tlfull{{\rm 2L\,diag\,full}}
\def\ctQEDfull{{\rm ct,QED\,full}}
\def\ctSUSYfull{{\rm ct,SUSY\,full}}
\def\ctrem{{\rm ct, rem}}

\newcommand{\fmfvcenter}[1]{\vcenter{\hbox{\fmfreuse{#1}}}}

\def\fmfct#1{\fmfv{label={\scalebox{1.2}{\large\boldmath{$\times$}}},label.dist=0}{#1}}
\def\fmfrct#1{\fmfv{label={\scalebox{1.2}{\rotatebox[origin=c]{33}{
                        \large\boldmath{$\times$\hspace{.5ex}}}}},label.dist=0}{#1}}

\begin{document}
\begin{flushright}
{\tt hep-ph/yymmnnn}\\
\end{flushright}
\vspace{3em}
\begin{center}
{\Large\bf Photonic SUSY Two-Loop Corrections\\[1ex]
 to the Muon Magnetic Moment}
\\
\vspace{3em}
{\sc P.\ von Weitershausen$^a$, M.\ Sch\"afer$^b$,
H.\ St\"ockinger-Kim$^c$,\\  D. St\"ockinger$^c$ 
}\\[2em]
{\sl ${}^a$Department of Physics and Astronomy, University of Glasgow,
Glasgow, UK}\\
{\sl ${}^b$Theoretisch-Physikalisches Institut, Universit\"at Jena,
Jena, Germany}\\
{\sl ${}^c$Institut f\"ur Kern- und Teilchenphysik,
TU Dresden, Dresden, Germany}
\setcounter{footnote}{0}
\end{center}
\vspace{2ex}
\begin{abstract}
Photonic SUSY two-loop corrections to the muon magnetic moment are
contributions from diagrams where an additional photon
loop is attached to a SUSY one-loop diagram. These photonic corrections are evaluated exactly, extending a
leading-log calculation by Degrassi and Giudice. 
Compact analytical expressions are provided and the numerical behaviour
is discussed. The photonic corrections reduce the
SUSY one-loop result by $7\ldots9\%$. The new terms are
typically around ten times smaller than the leading logarithms, but
they can be larger and have either sign in cases with large SUSY
mass splittings. We also provide details on renormalization and
regularization and on how to incorporate the photonic corrections into
a full SUSY two-loop calculation.
\end{abstract}

\vspace{0.5cm}
\centerline

\newpage

\section{Introduction}
\label{sec:introduction}

The muon magnetic dipole moment belongs to the most precisely known
observables in particle physics. The anomalous magnetic moment
$a_\mu=(g-2)_\mu/2$ has been determined by an impressive series of
measurements at BNL to \cite{ds-Bennett:2006}\footnote{%
The change in the number compared to Ref.\ \cite{ds-Bennett:2006} is
due to a new PDG value for the magnetic moment ratio of the muon to
proton \cite{Roberts:2010cj}.}
\begin{align}
a_\mu^{\rm exp} &= (11\,659\,208.9 \pm 6.3)\times 10^{-10} \, .
\end{align}
At this level of precision, $a_\mu$ is sensitive to all interactions
of the Standard Model (SM) as well as to hypothetical new particles at
the electroweak scale.

In recent years, the precision of the Standard Model (SM) theory
evaluation has reached a similar level. The hadronic vacuum
polarization contributions can be related to the cross section for
$e^+e^-\to$~hadrons in a  theoretically clean way, and  crucial recent
progress has been achieved on the experimental 
determination of this cross section by the SND, CMD-2, KLOE, and
Babar experiments \cite{SND,CMD2,KLOE,Babar}. The quality
of the current $e^+e^-\to$~hadrons data is such that all recent theory
evaluations of the corresponding contributions to $a_\mu$ agree 
fairly well, see
e.g.\ \cite{Hagiwara:2006jt,Teubner:2010ah,ds-deRafael:2008,JegerlehnerNyffeler,Davier:2009ag,Davier:2009zi}.\footnote{In
  principle, part of the $e^+e^-\to$~hadrons cross section could be 
obtained in an alternative way from hadronic $\tau$ decays
\cite{Alemany}. However, the required isospin breaking effects are difficult to
quantify at the currently needed level of precision
\cite{JegerlehnerGhozzi,MelnikovVainshtein,JegerlehnerNyffeler,Benayoun}.
Hence, in order to obtain the most reliable results, most analyses do
not use $\tau$ decays. Nevertheless, 
the most recent analysis \cite{Davier:2009ag} employing an improved
understanding of isospin breaking effects shows a marginal consistency
between the $e^+e^-$-based and the $\tau$-decay based evaluations of
$a_\mu$.}

The hadronic light-by-light contributions have been 
scrutinized by many groups in the past 15 years, and while an
ever better understanding has been achieved, the central value of the
result has remained relatively stable. Recently, three groups have
joined forces and published a common value, $a_\mu^{\rm
lbl}=10.5(2.6)\times10^{-10}$ \cite{dRPV}, where the errors have been
enlarged in order to cover the results obtained from the different
approaches.  Compatible results have been obtained even more recently
in Refs.\ \cite{JegerlehnerNyffeler,Nyffeler}. 
 
As a result of the recent progress, the SM theory prediction for
$a_\mu$ now has an even smaller error than $a_\mu^{\rm exp}$, calling
for a new experiment. For
reference we use the value from \cite{Davier:2009zi},
\begin{align}
a_{\mu}^{\rm SM}&= (11\,659\,183.4 \pm 4.9)\times 10^{-10}\, .
\end{align}
The deviation from the experimental value is
\begin{align}
\Delta a_{\mu}({\rm exp-SM}) &= (25.5 \pm 8.0 )  \times 10^{-10}\,
.\label{deviation} 
\end{align}
All the other mentioned recent  $e^+e^-$-based
evaluations lead to similar deviations in the range
$(25.5\ldots31.6)\times10^{-10}$ with combined errors in the range
$(7.9\ldots9.0)\times10^{-10}$.  

This  $3$--$4\sigma$ deviation constitutes a  tantalizing hint for physics
beyond the SM. Although the deviation is almost twice as large as the
SM weak contributions $a_\mu^{\rm weak}$ and contributions from
hypothetical heavy particles with mass $M$ are typically suppressed $\propto
(M_W/M)^2\,a_\mu^{\rm weak}$, there is a variety of models that could
explain it \cite{CzM}. Supersymmetry (SUSY) is a particularly promising
example (for a review see \cite{review}), owing to an enhancement by
$\tan\beta$, the ratio of the two Higgs vacuum expectation values. 

Even more importantly, the precision of the deviation
(\ref{deviation}) implies significant constraints on the parameters of many
new physics models. In supersymmetry, $a_\mu$ permits to derive mass
limits can which cannot be obtained from other observables
\cite{MaWe2}; $a_\mu$ is also one of the most important quantities in 
recent global analyses of the  parameter space of various
supersymmetric models \cite{Rosz,Allanach,Buchmueller,Fittino}. Even in the
LHC era, $a_\mu$ will remain a highly useful complementary
observable. It will provide a benchmark for models, help selecting
between different SUSY scenarios, eliminate ambiguities, and it will
improve parameter determinations. For instance,  combining
LHC data with $a_\mu$ will significantly reduce the $\tan\beta$
uncertainty \cite{WhitePaper,Sfitter,PlehnRauchNew}. 

Given the current situation and the usefulness of $a_\mu$, every
effort should be made in order to improve the experimental and
theoretical precision of $a_\mu$. 
In the near future, the precision of $\Delta a_{\mu}({\rm exp-SM})$
will indeed further increase. New analyses of
$e^+e^-\to$~hadrons data are in the pipeline, and a new CMD-3 experiment in
Novosibirsk is planned. All of these will directly feed into a more
precise evaluation of the hadronic vacuum polarization
contributions. Furthermore, a new, improved measurement of $a_\mu$
itself could be carried out at Fermilab \cite{FNALProposal} or
possibly at J-PARC \cite{Mibe}. Building
on the BNL experiment, whose final uncertainty is still statistics
dominated, the Fermilab measurement could reach a reduction of the
uncertainty down to $1.6\times10^{-10}$.

With these expected improvements, the theory error from contributions
from physics beyond the SM, in particular from supersymmetry,
becomes more prominent. The current theory error
of the SUSY contributions has been estimated to $3\times10^{-10}$
\cite{review}, which is larger than the precision goal of the Fermilab
$a_\mu$ measurement.

The  one-loop contributions from supersymmetric
particles have been computed a long time ago \cite{oneloop,moroi}, and
the SUSY two-loop corrections to SM one-loop diagrams (e.g.\
SM-diagrams with insertions of closed chargino/neutralino or
stop/sbottom loops) are completely known
\cite{HSW03,HSW04,ds-Feng}. However, the two-loop corrections to SUSY
one-loop diagrams are not known, only two dominant parts have been
identified: large QED-logarithms \cite{DG98} and $(\tan\beta)^2$-enhanced
corrections \cite{Marchetti:2008hw}. A subclass of the remaining
diagrams has been considered in Ref.\ \cite{FengLM06}. The unknown
two-loop contributions lead to the theory error mentioned above.

In the present paper we consider the photonic two-loop contributions,
defined as contributions from diagrams where a photon loop is attached
to a SUSY one-loop diagram. In Ref.\ \cite{DG98}, Degrassi and Giudice 
showed that these contributions lead to large QED-logarithms of the
form $\log(M/m_\mu)$, where $M$ is the mass scale of the new
particles. They evaluated the logarithms using elegant effective field 
theory and renormalization group techniques. These logarithms reduce
the one-loop contributions by $7\ldots9\%$ for $M$ between 100 and
1000 GeV.  It is clear that there is an ambiguity in the choice of
$M$, in particular in the case of a rather split spectrum. In the
present paper we evaluate the photonic two-loop
corrections exactly. In this way, we resolve the ambiguity, and we
obtain all logarithms of ratios of different heavy masses and the
associated non-logarithmic terms.
Our results are derived in a generic model which covers the case of
the minimal supersymmetric SM (MSSM) but also of a wider class of
models. We take into account issues such as the choice of dimensional 
regularization versus dimensional reduction, and we set up the
calculation in such a way that our results will be useful  building
blocks for a full two-loop 
computation of $a_\mu^{\rm SUSY}$. 

The outline of the paper is as
follows. In section \ref{sec:oneloop} we describe the setup of our
calculation and provide the necessary one-loop
contributions up to the order $\epsilon$ on the regularized level. In
section \ref{sec:twoloop} we list the contributing two-loop diagrams,
classify them according to their ultraviolet (UV) and infrared (IR)
divergences, and explain our method to compute them. Section
\ref{sec:cts} is devoted to renormalization and the analysis of the
counterterm diagrams. We provide details on the cancellation of UV and
IR divergences and the regularization-scheme dependence. 
 In section
\ref{sec:results} we provide our results as a compact analytical
formula, and discuss several numerical examples.
In section \ref{sec:matching} we discuss how our results can be used
as building blocks in a full MSSM two-loop calculation. Section
\ref{sec:conclusions} contains the conclusions.

\section{Generic model and one-loop contributions}
\label{sec:oneloop}

\begin{figure}[t]
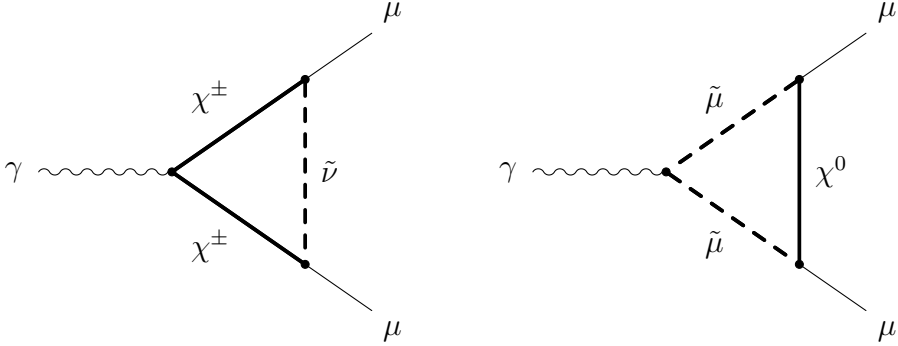

$$  \fmfvcenter{chaoneloop} \qquad  \qquad 
  \fmfvcenter{neuoneloop}$$
\caption{\label{fig:oneloop}
The two one-loop diagrams involving SUSY particles. In this and all
following figures, thin lines denote
the photon and muon, thick lines denote the chargino/neutralino, thick
dashed lines the sneutrino/smuon.
  }
\end{figure}

The genuine SUSY one-loop contributions to $a_\mu$ in SUSY extensions
of the SM are given by the two
kinds of diagrams in Fig.\ \ref{fig:oneloop}. The loops involve either
a chargino  $\chi^{\pm}$ and a sneutrino
$\tilde{\nu}_\mu$ or a neutralino $\chi^0$ and a smuon
$\tilde{\mu}$. At the two-loop level it is useful to classify the
SUSY contributions into those from diagrams where the $\mu$-lepton
number is only carried by muon or muon-neutrino, and diagrams where
the $\mu$-lepton number is carried also by $\tilde{\mu}$ and/or
$\tilde{\nu}_\mu$. The second class can be interpreted
as the two-loop corrections to the SUSY one-loop diagrams in
Fig.\ \ref{fig:oneloop}.

\newsavebox{\oneloopdiag}
\sbox{\oneloopdiag}{
\begin{fmffile}{feynman/oneloopdiag}
  \fmfset{thin}{.5pt}
  \fmfset{wiggly_len}{3mm}
  \fmfset{dash_len}{3.5mm}
  \fmfset{dot_size}{1.5thick}
  \fmfset{arrow_len}{2.5mm}

\begin{fmfgraph*}(140,105)
  \fmfkeep{chaoneloop}
  \fmfleft{gamma}
  \fmfright{in,out}
  \fmf{plain}{out,v1,va,vertex,vb,v2,in}
  \fmf{photon}{vertex,gamma}
  \fmf{dashes,width=1.5pt,tension=0,label=$\tilde{\nu}$,label.side=left}{v1,v2}
  \fmfdot{vertex,v2,v1}
\fmffreeze
\fmf{plain,width=1.5pt}{v1,va,vertex,vb,v2}
\fmf{plain,width=1.5pt,label=$\chi^{\pm}$,label.side=right}{v1,vertex,v2}
\fmflabel{$\gamma$}{gamma}
\fmflabel{$\mu$}{in}
\fmflabel{$\mu$}{out}
\end{fmfgraph*}

\begin{fmfgraph*}(140,105)
  \fmfkeep{neuoneloop}
  \fmfleft{gamma}
  \fmfright{in,out}
  \fmf{phantom}{out,v1,va,vertex,vb,v2,in}
  \fmf{photon}{vertex,gamma}
  \fmffreeze
  \fmf{plain}{out,v1,v2,in}
  \fmf{dashes,width=1.5pt}{v1,va,vertex,vb,v2}
  \fmf{dashes,width=1.5pt,label=$\tilde{\mu}$,label.side=right}{v1,vertex,v2}
  \fmfdot{vertex,v2,v1}
\fmffreeze
\fmf{plain,width=1.5pt,label=$\chi^{0}$,label.side=left}{v1,v2}
\fmflabel{$\gamma$}{gamma}
\fmflabel{$\mu$}{in}
\fmflabel{$\mu$}{out}
\end{fmfgraph*}
\end{fmffile}
}

In the present paper we aim for the computation of the photonic
contributions of this second class, i.e.\ of two-loop diagrams
obtained from attaching a photon loop in all possible ways to the SUSY
one-loop diagrams. The relevant Lagrangian is given by
\begin{align}
\label{model}
  {\cal L}_{{\rm QED}+c,n} &=
  {\cal L}_{\rm QED} + {\cal L}_{\rm mat} + {\cal L}_{\rm int},\\
{\cal L}_{\rm mat}&= \ 
  \overline{\chi^-} (i \Dsl - m_{\chi^\pm}) \chi^-
  + | D^\mu \tilde\mu |^2 - m_{\tilde\mu}^2 |\tilde\mu|^2 
\nonumber\\
  &+ \frac{1}{2}
  \overline{\chi^0} (i \partialsl - m_{\chi^0}) \chi^0
  + | \partial^\mu \tilde\nu |^2 - m_{\tilde\nu}^2 |\tilde\nu|^2
,\\
 {\cal L}_{\rm int} &= 
 \tilde\nu^\dagger \overline{ \chi^-} (c_L^* P_L + c_R P_R)\mu
+\tilde\mu^\dagger \overline{\chi^0}  (n_L^* P_L - n_R P_R)\mu
+ {\rm h.c.}
.
\end{align}
Here ${\cal L}_{\rm QED}$ is the QED Lagrangian for photon and muon $\mu$;
$\chi^-$ and $\chi^0$ are the chargino and neutralino spinor fields,
$D^\mu$ is the covariant derivative in QED and 
$P_{L,R}=\frac12(1\mp\gamma^5)$.
This Lagrangian contains only photon and muon plus one generic
chargino, neutralino, sneutrino, and smuon. For the following, it will
be useful to adopt the viewpoint that ${\cal L}_{{\rm QED}+c,n}$
describes a simple generic model which 
extends QED by two charged and two uncharged particles. The couplings
of the SUSY particles to the muon are given by free parameters
$c^{L,R}$ and $n^{L,R}$, all other terms in the Lagrangian are
determined by QED gauge invariance. 
The conventions of the couplings $c$,  $n$ are chosen in the same way
as in Ref.\ \cite{MartinWells}.
In terms of this generic model, the one-loop contributions
in Fig.\ \ref{fig:oneloop} and the photonic
two-loop corrections (plus renormalization) are defined as the
contributions of ${\cal O}(c^2,n^2)$ and ${\cal O}(\alpha c^2,\alpha
n^2)$,  
respectively, with the fine structure constant $\alpha=e^2/4\pi$.

By specializing the couplings, the generic model can describe photonic
corrections in underlying models such as the
MSSM, but also in non-minimal SUSY 
models, and even in non-supersymmetric models, if they happen to
contribute to $a_\mu$ by diagrams of the form in
Fig.\ \ref{fig:oneloop}. 
For instance, in the MSSM, the couplings $c^{L,R}$ and $n^{L,R}$ have
the following values:
\begin{subequations}
\label{cnrelations}
\begin{align}
c_{k}^L &= - g_2 V_{k1},\\
c_{k}^R &= y_\mu U_{k2},\\
n_{im}^L &= \frac{1}{\sqrt2}(g_1 N_{i1}+g_2 N_{i2}){U^{\tilde{\mu}}}_{m1}{}^*
-y_\mu N_{i3}{U^{\tilde{\mu}}}_{m2}{}^*,\\
n_{im}^R &= \sqrt2 g_1 N_{i1} {U^{\tilde{\mu}}}_{m2} + y_\mu
N_{i3}{U^{\tilde{\mu}}}_{m1},
\end{align}
\end{subequations}
where the chargino, neutralino and smuon indices can take the values
$k\in\{1,2\}$, $i\in\{1,\ldots,4\}$, $m\in\{1,2\}$, and where the
notation of Ref.\ \cite{review} has been used for gauge and Yukawa
couplings and mixing matrices. 


As a first step we present the result of the one-loop diagrams of
Fig.\ \ref{fig:oneloop}. These diagrams will be needed later as
counterterm diagrams, multiplied with $1/\epsilon$ divergences from
renormalization constants, where $D=4-2\epsilon$ in dimensional
regularization. Hence, we do not only need the finite part of the result
but also the ${\cal O}(\epsilon)$-part.

This ${\cal O}(\epsilon)$-part can in principle depend on the
regularization scheme. We consider two schemes: dimensional
regularization (DREG) and dimensional reduction (DRED). As discussed
e.g.\ in \cite{AS12}, it is crucial to define the treatment of
external vector bosons in dimensional schemes. Our definitions are as
follows: In DREG, the vertex function $\Gamma_{\mu\bar\mu A^\rho}$ is
computed with a $D$-dimensional photon, then the $D$-dimensional
projector to extract $a_\mu$ defined in \cite{CKProjOp,HSW04} is
applied. In DRED, the 
vertex function $\Gamma_{\mu\bar\mu A^\rho}$ is
computed with a (quasi-)4-dimensional photon, then the photon is
projected onto the $D$-dimensional subspace,
$\hat{g}_{\rho}{}^{\sigma}\Gamma_{\mu\bar\mu A^\sigma}$ in the notation of
\cite{AS12}, and finally the same $D$-dimensional $a_\mu$-projector
is applied.\footnote{Note that in DRED, for a quasi-4-dimensional
  photon, the covariant decomposition of $\Gamma_{\mu\bar\mu
    A^\sigma}$ has more terms than in DREG, and that different
  prescriptions to project out $a_\mu$ can lead to different results
  at ${\cal O}(\epsilon)$.}

With these definitions, it is clear that the regularized results for
the diagrams in Fig.\ \ref{fig:oneloop} are equal in DREG and DRED,
even at ${\cal O}(\epsilon)$. The results read
\begin{subequations}
\label{eqoneloop}
\begin{align}
{a_\mu^{\chi^\pm}}&=\frac{1}{16\pi^2}\frac{m_\mu^2}{m_{\tilde{\nu}}^2}
\Big\{
\frac{1}{12}\AA^C
{\cal F}_1^C(x)
+\frac{2m_{\chi^\pm}}{3}
\BB^C {\cal F}_2^C(x)
\Big\}
\label{amuchipm}
,\\
{a_\mu^{\chi^0}}&=\frac{1}{16\pi^2}\frac{m_\mu^2}{m_{\tilde{\mu}}^2}
\Big\{
-\frac{1}{12}\AA^N
{\cal F}_1^N(x)
+\frac{m_{\chi^0}}{3}
\BB^N {\cal F}_2^N(x)
\Big\}
,
\label{amuchi0}
\end{align}
\end{subequations}
with the kinematic variables
$x=m_{\chi^\pm}^2/m_{\tilde{\nu}_\mu}^2$, or
$x=m_{\chi^0}^2/m_{\tilde{\mu}}^2$, respectively, and the coupling
combinations
\begin{subequations}
\label{eq:AABBDef}
\begin{align}
\AA^C&=|c^L|^2+|c^R|^2
,&
\BB^C&=\frac{{\rm Re}[c^Lc^R]}{m_\mu}
,\\
\AA^N&=|n^L|^2+|n^R|^2
,&
\BB^N&=\frac{{\rm Re}[n^Ln^R]}{m_\mu}.
\end{align}
\end{subequations}
The $\epsilon$-dependent loop functions are decomposed as
\begin{align} 
\mathcal{F}^{C}_{i}(x) & = F^{C}_{i}(x)[1 - \epsilon \,
  \mathrm{L}(m_{\tilde{\nu}_{\mu}}^2)]  + 
\epsilon \, F^{C}_{i \epsilon}(x) ,\\ 
\mathcal{F}^{N}_{i}(x) & = F^{N}_{i}(x)[1 - \epsilon \,
  \mathrm{L}(m_{\tilde{\mu}}^2)]  + 
\epsilon \, F^{N}_{i \epsilon}(x) ,
\end{align}
where we have used the abbreviation
\begin{align}
{\rm L}(m^2) &= \log\frac{m^2}{\mu^2_{\rm DREG}}
\label{LDef}
\end{align}
with the dimensional-regularization scale $\mu_{\rm DREG}$,
and the well-known functions
\begin{align}
F_1^C(x) &= \frac{ 2 }{(1-x)^4 }\big[
2+3x-6x^2+x^3+6x\log x
\big],\\
F_2^C(x) &= \frac{ 3 }{2(1-x)^3 }\big[
-3+4x-x^2-2\log x
\big],\\
F_1^N(x) &= \frac{ 2 }{(1-x)^4 }\big[
1-6x+3x^2+2x^3-6x^2\log x
\big],\\
F_2^N(x) &= \frac{ 3 }{(1-x)^3 }\big[
1-x^2+2x\log x
\big],
\end{align}
normalized such that $F_i^j(1)=1$. The functions for the
$\epsilon$-dependent parts are defined as 
\begin{align}
F^{C}_{1\epsilon}(x) & = F^{C}_{1}(x) \left(
\frac{-x^3 + 6 x^2 + 15 x + 2 - 6 x \log x}{12 x} \right) +
\frac{x^2 - 8 x - 4}{6 x}
 \label{loopeqfceps1}
,\\
F^{C}_{2\epsilon}(x) & = F^{C}_{2}(x)  \left(
\frac{-2 x^2 + 8 x + 6 - 4 \log x}{8} \right) + \frac{3 x -
  15}{8} \label{loopeqfceps2}
,\\
F^{N}_{1\epsilon}(x) & = F^{N}_{1}(x) \left( \frac{2
  x^3 +  15 x^2 + 6 x - 1 - 6 x ^2 \log x}{ 12 x^2} \right) + \frac{1
  - 8 x - 4 x^2}{6 x^2}, \label{loopeqfneps1}\\ 
F^{N}_{2\epsilon}(x) & = F^{N}_{2}(x) \left(
\frac{x^2 + 4 x + 1 - 2 x \log x}{4 x} \right) - \frac{3 x + 3}{4 x},
 \label{loopeqfneps2}
\end{align}
and are normalized to $F^j_{i\epsilon}(1)=0$.
The ${\cal O}(\epsilon^0)$-part of
this result has been presented in a form similar to the one in
\cite{MartinWells}.

Our definition of the coupling combinations $\AA$, $\BB$ reflects an
important physical property.  The muon anomalous magnetic moment, like
the muon mass, corresponds to a chirality-flipping operator, and all
contributions must be proportional to a chiral-symmetry violating
parameter. In the SM, the MSSM and many other models, chiral symmetry
in the muon sector is broken only by the muon mass and its Yukawa
coupling. In the remainder of the article we assume nothing about the
model underlying the Lagrangian (\ref{model}) except that it has this
same property. 

With this assumption, all contributions to $a_\mu$ are of
the order $m_\mu^2/M_{\rm SUSY}^2$ in the muon mass, like in the MSSM
\cite{moroi,CzM,review}. Furthermore, the combinations $\AA^{C,N}$ and
$\BB^{C,N}$ must have the behaviour
\begin{align}
\AA^{C,N},\BB^{C,N} &= {\rm const.}+{\cal O}(m^2_\mu)
\end{align}
in the muon mass. In other words, either $c^L$ or $c^R$ must be
proportional to the muon mass, and similar for $n^L$, $n^R$.
It is noteworthy that an equivalent assumption has been made
implicitly in Ref.\ \cite{DG98} in the computation of the 2-loop
QED-logarithms in new physics models using effective field theory
techniques. In the matching between the full and the effective theory
Ref.\ \cite{DG98} assumed that one can pull a factor $m_\mu^2$ out of
the new physics contributions to $a_\mu$. If that were not the case,
additional large QED-logarithms would appear in the matching.

For the case of the MSSM, Eq.\ (\ref{cnrelations})
shows that the couplings indeed have the assumed behaviour, for all
values of the chargino, neutralino and smuon 
indices\footnote{In 
  order to see this, note that the muon Yukawa coupling $y_\mu$ and,
  for each $m$, one of the mixing matrix elements
  ${U^{\tilde{\mu}}}_{m1}$ or ${U^{\tilde{\mu}}}_{m2}$ are
  proportional to $m_\mu$.}.
For phenomenology it is important that $\BB^{C,N}$ are enhanced by a
factor $\tan\beta$. In a large part of the MSSM parameter space, the
$\BB^C$-term in $a_\mu^{\chi^\pm}$ constitutes the dominant SUSY
contribution to $a_\mu$. 


\section{Two-loop contributions}
\label{sec:twoloop}
\newsavebox{\chadiagrams}
\sbox{\chadiagrams}{
\begin{fmffile}{feynman/qedcha}
  \fmfset{thin}{.5pt}
  \fmfset{wiggly_len}{3mm}
  \fmfset{dash_len}{2.5mm}
  \fmfset{dot_size}{1.5thick}
  \fmfset{arrow_len}{2.5mm}

\begin{fmfgraph*}(80,60)
  \fmfkeep{qedcha-1a}
  \fmfleft{gamma}
  \fmfright{in,out}
  \fmf{plain}{out,va,v1,vb,vertex,vc,v2,vd,in}
  \fmf{photon}{vertex,gamma}
  \fmf{photon,tension=0}{v1,v2}
  \fmffreeze
  \fmf{dashes,width=1.5pt,right}{va,vb}
\fmf{plain,width=1.5pt}{va,v1,vb}
  \fmfdot{vertex,v1,v2,va,vb}
\end{fmfgraph*}

\begin{fmfgraph*}(80,60)
  \fmfkeep{qedcha-1b}
  \fmfleft{gamma}
  \fmfright{in,out}
  \fmf{plain}{out,va,v1,vb,vertex,vc,v2,vd,in}
  \fmf{photon}{vertex,gamma}
  \fmf{photon,tension=0}{v1,v2}
  \fmffreeze
  \fmf{dashes,width=1.5pt,right}{vc,vd}
\fmf{plain,width=1.5pt}{vc,v2,vd}
  \fmfdot{vertex,v1,v2,vc,vd}
\end{fmfgraph*}

\begin{fmfgraph*}(80,60)
  \fmfset{dash_len}{2mm} 
  \fmfkeep{qedcha-2a}
  \fmfleft{gamma}
  \fmfright{in,out}
  \fmf{plain}{out,v1,va}
  \fmf{plain,left,width=1.5pt}{va,vb}
  \fmf{plain}{vb,vertex,vc,vd,v2,in}
  \fmf{photon}{vertex,gamma}
  \fmf{photon,tension=0}{v1,v2}
  \fmffreeze
  \fmf{dashes,width=1.5pt,right}{va,vb}
  \fmfdot{vertex,v1,v2,va,vb}
\end{fmfgraph*}

\begin{fmfgraph*}(80,60)
  \fmfset{dash_len}{2mm} 
  \fmfkeep{qedcha-2b}
  \fmfleft{gamma}
  \fmfright{in,out}
  \fmf{plain}{out,v1,va,vb,vertex,vc}
  \fmf{plain,left,width=1.5pt}{vc,vd}
  \fmf{plain}{vd,v2,in}
  \fmf{photon}{vertex,gamma}
  \fmf{photon,tension=0}{v1,v2}
  \fmffreeze
  \fmf{dashes,width=1.5pt,right}{vc,vd}
  \fmfdot{vertex,v1,v2,vc,vd}
\end{fmfgraph*}

\begin{fmfgraph*}(80,60)
  \fmfkeep{qedcha-3a}
  \fmfleft{gamma}
  \fmfright{in,out}
  \fmf{plain}{out,va,v1,vb,vertex,vc,v2,vd,in}
  \fmf{photon}{vertex,gamma}
  \fmf{dashes,width=1.5pt,tension=0}{v1,v2}
  \fmffreeze
\fmf{plain,width=1.5pt}{v1,vb,vertex,vc,v2}
  \fmf{photon,right}{va,vb}
  \fmfdot{vertex,v1,v2,va,vb}
\end{fmfgraph*}

\begin{fmfgraph*}(80,60)
  \fmfkeep{qedcha-3b}
  \fmfleft{gamma}
  \fmfright{in,out}
  \fmf{plain}{out,va,v1,vb,vertex,vc,v2,vd,in}
  \fmf{photon}{vertex,gamma}
  \fmf{dashes,width=1.5pt,tension=0}{v1,v2}
  \fmffreeze
\fmf{plain,width=1.5pt}{v1,vb,vertex,vc,v2}
  \fmf{photon,right}{vc,vd}
  \fmfdot{vertex,v1,v2,vc,vd}
\end{fmfgraph*}

\begin{fmfgraph*}(80,60)
  \fmfkeep{qedcha-4a}
  \fmfleft{gamma}
  \fmfright{in,out}
  \fmf{plain}{out,va,v1,vertex,v2,vb,in}
  \fmf{photon}{vertex,gamma}
  \fmffreeze
\fmf{plain,width=1.5pt}{v1,vertex,v2,vb}
  \fmf{dashes,width=1.5pt}{v1,vb}
  \fmf{photon}{v2,va}
  \fmfdot{vertex,v1,v2,va,vb}
\end{fmfgraph*}

\begin{fmfgraph*}(80,60)
  \fmfkeep{qedcha-4b}
  \fmfleft{gamma}
  \fmfright{in,out}
  \fmf{plain}{out,va,v1,vertex,v2,vb,in}
  \fmf{photon}{vertex,gamma}
  \fmffreeze
  \fmf{dashes,width=1.5pt}{v2,va}
\fmf{plain,width=1.5pt}{va,v1,vertex,v2}
  \fmf{photon}{v1,vb}
  \fmfdot{vertex,v1,v2,va,vb}
\end{fmfgraph*}

\begin{fmfgraph*}(80,60)
  \fmfkeep{qedcha-5}
  \fmfleft{gamma}
  \fmfright{in,out}
  \fmf{plain}{out,va,v1,vertex,v2,vb,in}
  \fmf{photon}{vertex,gamma}
  \fmffreeze
  \fmf{dashes,width=1.5pt}{v1,v2}
\fmf{plain,width=1.5pt}{v1,vertex,v2}
  \fmf{photon}{va,vb}
  \fmfdot{vertex,v1,v2,va,vb}
\end{fmfgraph*}

\begin{fmfgraph*}(80,60)
  \fmfkeep{qedcha-6a}
  \fmfleft{gamma}
  \fmfright{in,out}
  \fmf{plain}{out,v1,va,vb,vertex,vc,vd,v2,in}
  \fmf{photon}{vertex,gamma}
  \fmf{dashes,width=1.5pt,tension=0}{v1,v2}
  \fmffreeze
\fmf{plain,width=1.5pt}{v1,va,vb,vertex,vc,vd,v2}
  \fmf{photon,right}{va,vb}
  \fmfdot{vertex,v1,v2,va,vb}
\end{fmfgraph*}

\begin{fmfgraph*}(80,60)
  \fmfkeep{qedcha-6b}
  \fmfleft{gamma}
  \fmfright{in,out}
  \fmf{plain}{out,v1,va,vb,vertex,vc,vd,v2,in}
  \fmf{photon}{vertex,gamma}
  \fmf{dashes,width=1.5pt,tension=0}{v1,v2}
  \fmffreeze
\fmf{plain,width=1.5pt}{v1,va,vb,vertex,vc,vd,v2}
  \fmf{photon,right}{vc,vd}
  \fmfdot{vertex,v1,v2,vc,vd}
\end{fmfgraph*}

\begin{fmfgraph*}(80,60)
  \fmfkeep{qedcha-7}
  \fmfleft{gamma}
  \fmfright{in,out}
  \fmf{plain}{out,va,v1,vertex,v2,vb,in}
  \fmf{photon}{vertex,gamma}
  \fmffreeze
  \fmf{photon}{v1,v2}
  \fmf{dashes,width=1.5pt}{va,vb}
\fmf{plain,width=1.5pt}{va,v1,vertex,v2,vb}
  \fmfdot{vertex,v1,v2,va,vb}
\end{fmfgraph*}
\end{fmffile}
}

\newsavebox{\neudiagrams}
\sbox{\neudiagrams}{
\begin{fmffile}{feynman/qedneu}
  \fmfset{thin}{.5pt}
  \fmfset{wiggly_len}{3mm}
  \fmfset{dash_len}{2.5mm}
  \fmfset{dot_size}{1.5thick}
  \fmfset{arrow_len}{2.5mm}

\begin{fmfgraph*}(80,60)
  \fmfkeep{qedneu-1a}
  \fmfleft{gamma}
  \fmfright{in,out}
  \fmf{plain}{out,va}
  \fmf{dashes,width=1.5pt}{va,v1,vb}
  \fmf{plain}{vb,vertex,vc,v2,vd,in}
  \fmf{photon}{vertex,gamma}
  \fmf{photon,tension=0}{v1,v2}
  \fmffreeze
  \fmf{plain,width=1.5pt,right}{va,vb}
  \fmfdot{vertex,v1,v2,va,vb}
\end{fmfgraph*}

\begin{fmfgraph*}(80,60)
  \fmfkeep{qedneu-1b}
  \fmfleft{gamma}
  \fmfright{in,out}
  \fmf{plain}{out,va,v1,vb,vertex,vc}
  \fmf{dashes,width=1.5pt}{vc,v2,vd}
  \fmf{plain}{vd,in}
  \fmf{photon}{vertex,gamma}
  \fmf{photon,tension=0}{v1,v2}
  \fmffreeze
  \fmf{plain,width=1.5pt,right}{vc,vd}
  \fmfdot{vertex,v1,v2,vc,vd}
\end{fmfgraph*}

\begin{fmfgraph*}(80,60)
  \fmfset{dash_len}{2mm} 
  \fmfkeep{qedneu-2a}
  \fmfleft{gamma}
  \fmfright{in,out}
  \fmf{plain}{out,v1,va}
  \fmf{dashes,width=1.5pt,left}{va,vb}
  \fmf{plain}{vb,vertex,vc,vd,v2,in}
  \fmf{photon}{vertex,gamma}
  \fmf{photon,tension=0}{v1,v2}
  \fmffreeze
  \fmf{plain,width=1.5pt,right}{va,vb}
  \fmfdot{vertex,v1,v2,va,vb}
\end{fmfgraph*}

\begin{fmfgraph*}(80,60)
  \fmfset{dash_len}{2mm} 
  \fmfkeep{qedneu-2b}
  \fmfleft{gamma}
  \fmfright{in,out}
  \fmf{plain}{out,v1,va,vb,vertex,vc}
  \fmf{dashes,width=1.5pt,left}{vc,vd}
  \fmf{plain}{vd,v2,in}
  \fmf{photon}{vertex,gamma}
  \fmf{photon,tension=0}{v1,v2}
  \fmffreeze
  \fmf{plain,width=1.5pt,right}{vc,vd}
  \fmfdot{vertex,v1,v2,vc,vd}
\end{fmfgraph*}

\begin{fmfgraph*}(80,60)
  \fmfkeep{qedneu-3a}
  \fmfleft{gamma}
  \fmfright{in,out}
  \fmf{phantom}{out,va,v1,vb,vertex,vc,v2,vd,in}
  \fmf{plain,tension=0}{out,va,v1,v2,vd,in}
  \fmf{photon}{vertex,gamma}
  \fmffreeze
\fmf{plain,width=1.5pt}{v1,v2}
  \fmf{dashes,width=1.5pt}{v1,vb,vertex,vc,v2}
  \fmf{photon,right}{va,vb}
  \fmfdot{vertex,v1,v2,va,vb}
\end{fmfgraph*}

\begin{fmfgraph*}(80,60)
  \fmfkeep{qedneu-3b}
  \fmfleft{gamma}
  \fmfright{in,out}
  \fmf{phantom}{out,va,v1,vb,vertex,vc,v2,vd,in}
  \fmf{plain,tension=0}{out,va,v1,v2,vd,in}
  \fmf{photon}{vertex,gamma}
  \fmffreeze
\fmf{plain,width=1.5pt}{v1,v2}
  \fmf{dashes,width=1.5pt}{v1,vb,vertex,vc,v2}
  \fmf{photon,right}{vc,vd}
  \fmfdot{vertex,v1,v2,vc,vd}
\end{fmfgraph*}

\begin{fmfgraph*}(80,60)
  \fmfkeep{qedneu-3'a}
  \fmfleft{gamma}
  \fmfright{in,out}
  \fmf{phantom}{out,va,v1,vertex,v2,vd,in}
  \fmf{plain,tension=0}{out,va,v1,v2,vd,in}
  \fmf{photon}{vertex,gamma}
  \fmffreeze
\fmf{plain,width=1.5pt}{v1,v2}
  \fmf{dashes,width=1.5pt}{v1,vertex,v2}
  \fmf{photon,right}{va,vertex}
  \fmfdot{vertex,v1,v2,va}
\end{fmfgraph*}

\begin{fmfgraph*}(80,60)
  \fmfkeep{qedneu-3'b}
  \fmfleft{gamma}
  \fmfright{in,out}
  \fmf{phantom}{out,va,v1,vertex,v2,vd,in}
  \fmf{plain,tension=0}{out,va,v1,v2,vd,in}
  \fmf{photon}{vertex,gamma}
  \fmffreeze
\fmf{plain,width=1.5pt}{v1,v2}
  \fmf{dashes,width=1.5pt}{v1,vertex,v2}
  \fmf{photon,right}{vertex,vd}
  \fmfdot{vertex,v1,v2,vd}
\end{fmfgraph*}

\begin{fmfgraph*}(80,60)
  \fmfkeep{qedneu-4a}
  \fmfleft{gamma}
  \fmfright{in,out}
  \fmf{phantom}{out,va,v1,vertex,v2,vb,in}
  \fmf{plain,tension=0}{out,va,v1,vb,in}
  \fmf{photon}{vertex,gamma}
  \fmffreeze
\fmf{plain,width=1.5pt}{v1,vb}
  \fmf{dashes,width=1.5pt}{v1,vertex,v2,vb}
  \fmf{photon}{v2,va}
  \fmfdot{vertex,v1,v2,va,vb}
\end{fmfgraph*}

\begin{fmfgraph*}(80,60)
  \fmfkeep{qedneu-4b}
  \fmfleft{gamma}
  \fmfright{in,out}
  \fmf{phantom}{out,va,v1,vertex,v2,vb,in}
  \fmf{plain,tension=0}{out,va,v2,vb,in}
  \fmf{photon}{vertex,gamma}
  \fmffreeze
\fmf{plain,width=1.5pt}{va,v2}
  \fmf{dashes,width=1.5pt}{va,v1,vertex,v2}
  \fmf{photon}{v1,vb}
  \fmfdot{vertex,v1,v2,va,vb}
\end{fmfgraph*}

\begin{fmfgraph*}(80,60)
  \fmfkeep{qedneu-5}
  \fmfleft{gamma}
  \fmfright{in,out}
  \fmf{phantom}{out,va,v1,vertex,v2,vb,in}
  \fmf{plain,tension=0}{out,va,v1,v2,vb,in}
  \fmf{photon}{vertex,gamma}
  \fmffreeze
\fmf{plain,width=1.5pt}{v1,v2}
  \fmf{dashes,width=1.5pt}{v1,vertex,v2}
  \fmf{photon}{va,vb}
  \fmfdot{vertex,v1,v2,va,vb}
\end{fmfgraph*}

\begin{fmfgraph*}(80,60)
  \fmfkeep{qedneu-6a}
  \fmfleft{gamma}
  \fmfright{in,out}
  \fmf{phantom}{out,v1,va,vb,vertex,vc,vd,v2,in}
  \fmf{plain,tension=0}{out,v1,v2,in}
  \fmf{photon}{vertex,gamma}
  \fmffreeze
\fmf{plain,width=1.5pt}{v1,v2}
  \fmf{dashes,width=1.5pt}{v1,va,vb,vertex,vc,vd,v2}
  \fmf{photon,right}{va,vb}
  \fmfdot{vertex,v1,v2,va,vb}
\end{fmfgraph*}

\begin{fmfgraph*}(80,60)
  \fmfkeep{qedneu-6b}
  \fmfleft{gamma}
  \fmfright{in,out}
  \fmf{phantom}{out,v1,va,vb,vertex,vc,vd,v2,in}
  \fmf{plain,tension=0}{out,v1,v2,in}
  \fmf{photon}{vertex,gamma}
  \fmffreeze
\fmf{plain,width=1.5pt}{v1,v2}
  \fmf{dashes,width=1.5pt}{v1,va,vb,vertex,vc,vd,v2}
  \fmf{photon,right}{vc,vd}
  \fmfdot{vertex,v1,v2,vc,vd}
\end{fmfgraph*}

\begin{fmfgraph*}(80,60)
  \fmfkeep{qedneu-6'a}
  \fmfleft{gamma}
  \fmfright{in,out}
  \fmf{phantom}{out,v1,va,vertex,vd,v2,in}
  \fmf{plain,tension=0}{out,v1,v2,in}
  \fmf{photon}{vertex,gamma}
  \fmffreeze
\fmf{plain,width=1.5pt}{v1,v2}
  \fmf{dashes,width=1.5pt}{v1,va,vertex,vd,v2}
  \fmf{photon,right}{va,vertex}
  \fmfdot{vertex,v1,v2,va}
\end{fmfgraph*}

\begin{fmfgraph*}(80,60)
  \fmfkeep{qedneu-6'b}
  \fmfleft{gamma}
  \fmfright{in,out}
  \fmf{phantom}{out,v1,va,vertex,vd,v2,in}
  \fmf{plain,tension=0}{out,v1,v2,in}
  \fmf{photon}{vertex,gamma}
  \fmffreeze
\fmf{plain,width=1.5pt}{v1,v2}
  \fmf{dashes,width=1.5pt}{v1,va,vertex,vd,v2}
  \fmf{photon,right}{vertex,vd}
  \fmfdot{vertex,v1,v2,vd}
\end{fmfgraph*}

\begin{fmfgraph*}(80,60)
  \fmfkeep{qedneu-7}
  \fmfleft{gamma}
  \fmfright{in,out}
  \fmf{phantom}{out,va,v1,vertex,v2,vb,in}
  \fmf{plain,tension=0}{out,va,vb,in}
  \fmf{photon}{vertex,gamma}
  \fmffreeze
\fmf{plain,width=1.5pt}{va,vb}
  \fmf{dashes,width=1.5pt}{va,v1,vertex,v2,vb}
  \fmf{photon}{v1,v2}
  \fmfdot{vertex,v1,v2,va,vb}
\end{fmfgraph*}
\end{fmffile}
}

\begin{figure}
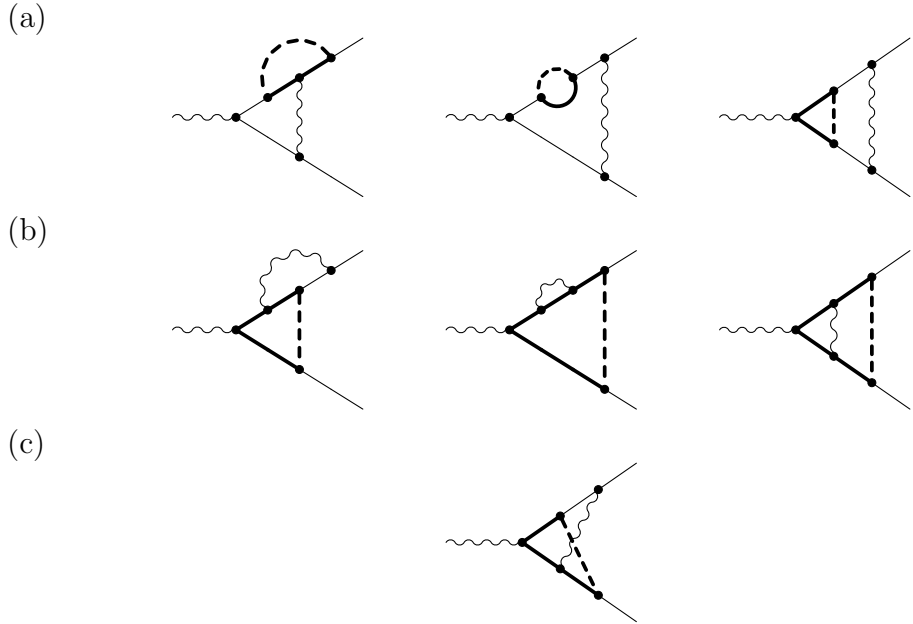

(a) 
$$ \fmfvcenter{qedcha-1a} \qquad
   \fmfvcenter{qedcha-2a} \qquad
   \fmfvcenter{qedcha-5}$$
(b)
$$  \fmfvcenter{qedcha-3a} \qquad
    \fmfvcenter{qedcha-6a} \qquad
    \fmfvcenter{qedcha-7}$$
(c)
$$  \fmfvcenter{qedcha-4a}$$
\caption{\label{fig:twoloopCha} Photonic two-loop corrections to
  chargino one-loop diagrams. In this and all following figures, it is
understood that many graphs (here all graphs except the 3rd and 6th)
appear twice, with the SUSY loop either on the upper or on the lower
muon line. These graphs are displayed only once.}
\end{figure}

\begin{figure}
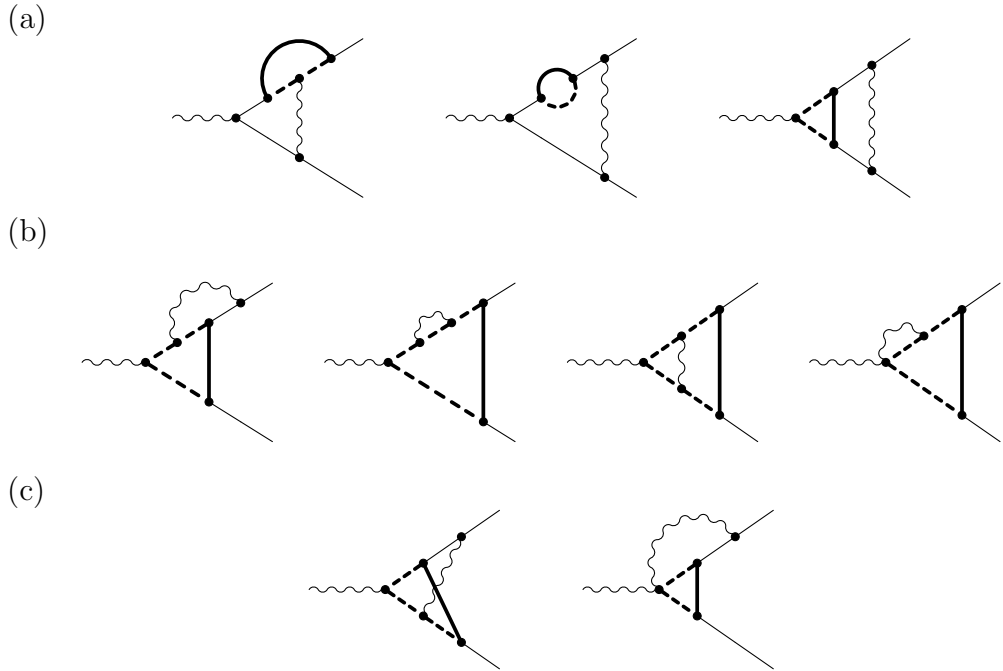

(a)
$$   \fmfvcenter{qedneu-1a} \qquad
     \fmfvcenter{qedneu-2a} \qquad
     \fmfvcenter{qedneu-5}  $$
(b)
$$  \fmfvcenter{qedneu-3a} \quad
    \fmfvcenter{qedneu-6a} \quad
    \fmfvcenter{qedneu-7}  \quad
    \fmfvcenter{qedneu-6'a}   $$
(c)
$$   \fmfvcenter{qedneu-4a} \qquad
     \fmfvcenter{qedneu-3'a} $$
\caption{\label{fig:twoloopNeu} Photonic two-loop corrections to
  neutralino one-loop diagrams.}
\end{figure}
We are interested in the photonic two-loop corrections to
Fig.\ \ref{fig:oneloop}, i.e.\ the two-loop
contributions of ${\cal O}(\alpha c^2,\alpha n^2)$ in our generic
model. We denote the corresponding genuine two-loop diagrams by
$a_\mu^{\tlgenuine}$. They are shown in
Figs.\ \ref{fig:twoloopCha}, \ref{fig:twoloopNeu}, divided into
diagrams with chargino/sneutrino or neutralino/smuon exchange. All
diagrams are overall UV-convergent, but a useful classification
according to divergent subdiagrams is possible. 
\begin{itemize}
\item The diagrams in
Figs.\  \ref{fig:twoloopCha}a and \ref{fig:twoloopNeu}a contain
a UV-divergent SUSY subdiagram, inserted into a QED diagram. 
\item The diagrams in Figs.\  \ref{fig:twoloopCha}b and
  \ref{fig:twoloopNeu}b contain 
a UV-divergent QED subdiagram, inserted into a SUSY diagram. There are
more neutralino/smuon diagrams because of the presence of vertices
involving two photons.
\item The diagrams in Figs.\  \ref{fig:twoloopCha}c and
  \ref{fig:twoloopNeu}c contain no UV-divergent subdiagrams and 
  are finite.
\end{itemize}

There are also IR divergences in the third diagrams in
Figs.\ \ref{fig:twoloopCha}a and \ref{fig:twoloopNeu}a, which are
the only diagrams where both ends of the photon propagator are
attached to the external muons and with a subdiagram that contributes
to $a_\mu$. 

We have evaluated the diagrams using DREG as a regulator for UV and IR
divergences. In the next section we will discuss how the UV and IR
divergences of the different classes of diagrams are cancelled by
renormalization, and we will discuss the transition to
DRED. Technically, we have employed computer algebra programs based on the
packages {\em FeynArts} \cite{feynarts} (using both a custom model
file and the standard MSSM model file \cite{fa-mssm})
and {\em TwoCalc} \cite{2lred} (with a custom routine for the integral
reduction). 

We have applied a large mass expansion \cite{smirnov} to all diagrams,
where the muon mass is treated as small and all other masses as
large. This results in a separation of heavy and light scales, and
after reduction to master integrals only two kinds of loop
integrals can appear: 
\begin{itemize}
\item products of a light one-loop integral (depending only on
  $m_\mu$) and a heavy one-loop integral (independent of $m_\mu$):
  these are the integrals in which the large $\log m_\mu$-factors
  considered in \cite{DG98} but also other terms are generated.
\item heavy two-loop integrals (independent of $m_\mu$, at most
  multiplied with polynomials in $m_\mu$): no large logarithms are
  generated here; these contributions are new. The heavy two-loop
  integrals have no external momentum and can depend only on two
  different mass scales.
\end{itemize}
The structure of these integrals makes clear that $\log m_\mu$-factors
appear in all diagrams in which the internal photon couples to at
least one muon. 
Furthermore, the diagrams of Figs.\ \ref{fig:twoloopCha}a and
\ref{fig:twoloopNeu}a contain terms of order $(m_\mu/M_{\rm SUSY})^0$,
while all other diagrams are suppressed by the small ratio
$(m_\mu/M_{\rm SUSY})^2$. 
The final result of all diagrams can be expressed
as a rational function of the particle masses, and logarithms and
dilogarithms. Below we will present the result obtained after
renormalization.

\section{Renormalization and counterterms}
\label{sec:cts}

\newsavebox{\qedcounterterms}
\sbox{\qedcounterterms}{
\begin{fmffile}{feynman/qedct}
  \fmfset{thin}{.5pt}
  \fmfset{wiggly_len}{3mm}
  \fmfset{dash_len}{2.5mm}
  \fmfset{dot_size}{1.5thick}
  \fmfset{arrow_len}{2.5mm}

\begin{fmfgraph*}(80,60)
  \fmfkeep{qedct-1a}
  \fmfleft{gamma}
  \fmfright{in,out}
  \fmf{plain}{out,v1,va,vertex,vb,v2,in}
  \fmf{photon}{vertex,gamma}
  \fmf{photon,tension=0}{v1,v2}
  \fmfdot{vertex,v2}
  \fmfct{v1}
\end{fmfgraph*}

\begin{fmfgraph*}(80,60)
  \fmfkeep{qedct-1b}
  \fmfleft{gamma}
  \fmfright{in,out}
  \fmf{plain}{out,v1,va,vertex,vb,v2,in}
  \fmf{photon}{vertex,gamma}
  \fmf{photon,tension=0}{v1,v2}
  \fmfdot{v1,vertex}
  \fmfct{v2}
\end{fmfgraph*}

\begin{fmfgraph*}(80,60)
  \fmfkeep{qedct-2a}
  \fmfleft{gamma}
  \fmfright{in,out}
  \fmf{plain}{out,v1,va,vertex,vb,v2,in}
  \fmf{photon}{vertex,gamma}
  \fmf{photon,tension=0}{v1,v2}
  \fmfdot{v1,vertex,v2}
  \fmfrct{va}
\end{fmfgraph*}

\begin{fmfgraph*}(80,60)
  \fmfkeep{qedct-2b}
  \fmfleft{gamma}
  \fmfright{in,out}
  \fmf{plain}{out,v1,va,vertex,vb,v2,in}
  \fmf{photon}{vertex,gamma}
  \fmf{photon,tension=0}{v1,v2}
  \fmfdot{v1,vertex,v2}
  \fmfrct{vb}
\end{fmfgraph*}

\begin{fmfgraph*}(80,60)
  \fmfkeep{qedct-5}
  \fmfleft{gamma}
  \fmfright{in,out}
  \fmf{plain}{out,v1,va,vertex,vb,v2,in}
  \fmf{photon}{vertex,gamma}
  \fmf{photon,tension=0}{v1,v2}
  \fmfdot{v1,v2}
  \fmfct{vertex}
\end{fmfgraph*}
\end{fmffile}
}

\newsavebox{\chacounterterms}
\sbox{\chacounterterms}{
\begin{fmffile}{feynman/qedchact}
  \fmfset{thin}{.5pt}
  \fmfset{wiggly_len}{3mm}
  \fmfset{dash_len}{2.5mm}
  \fmfset{dot_size}{1.5thick}
  \fmfset{arrow_len}{2.5mm}

\begin{fmfgraph*}(80,60)
  \fmfkeep{qedchact-3a}
  \fmfleft{gamma}
  \fmfright{in,out}
  \fmf{plain}{out,v1,va,vertex,vb,v2,in}
  \fmf{photon}{vertex,gamma}
  \fmf{dashes,width=1.5pt,tension=0}{v1,v2}
  \fmfdot{vertex,v2}
  \fmfct{v1}
\fmffreeze
\fmf{plain,width=1.5pt}{v1,va,vertex,vb,v2}
\end{fmfgraph*}

\begin{fmfgraph*}(80,60)
  \fmfkeep{qedchact-3b}
  \fmfleft{gamma}
  \fmfright{in,out}
  \fmf{plain}{out,v1,va,vertex,vb,v2,in}
  \fmf{photon}{vertex,gamma}
  \fmf{dashes,width=1.5pt,tension=0}{v1,v2}
  \fmfdot{v1,vertex}
  \fmfct{v2}
\fmffreeze
\fmf{plain,width=1.5pt}{v1,va,vertex,vb,v2}
\end{fmfgraph*}

\begin{fmfgraph*}(80,60)
  \fmfkeep{qedchact-6a}
  \fmfleft{gamma}
  \fmfright{in,out}
  \fmf{plain}{out,v1,va,vertex,vb,v2,in}
  \fmf{photon}{vertex,gamma}
  \fmf{dashes,width=1.5pt,tension=0}{v1,v2}
  \fmfdot{v1,vertex,v2}
  \fmfrct{va}
\fmffreeze
\fmf{plain,width=1.5pt}{v1,va,vertex,vb,v2}
\end{fmfgraph*}

\begin{fmfgraph*}(80,60)
  \fmfkeep{qedchact-6b}
  \fmfleft{gamma}
  \fmfright{in,out}
  \fmf{plain}{out,v1,va,vertex,vb,v2,in}
  \fmf{photon}{vertex,gamma}
  \fmf{dashes,width=1.5pt,tension=0}{v1,v2}
  \fmfdot{v1,vertex,v2}
  \fmfct{vb}
\fmffreeze
\fmf{plain,width=1.5pt}{v1,va,vertex,vb,v2}
\end{fmfgraph*}

\begin{fmfgraph*}(80,60)
  \fmfkeep{qedchact-7}
  \fmfleft{gamma}
  \fmfright{in,out}
  \fmf{plain}{out,v1,va,vertex,vb,v2,in}
  \fmf{photon}{vertex,gamma}
  \fmf{dashes,width=1.5pt,tension=0}{v1,v2}
  \fmfdot{v1,v2}
  \fmfct{vertex}
\fmffreeze
\fmf{plain,width=1.5pt}{v1,va,vertex,vb,v2}
\end{fmfgraph*}

\end{fmffile}
}

\newsavebox{\neucounterterms}
\sbox{\neucounterterms}{
\begin{fmffile}{feynman/qedneuct}
  \fmfset{thin}{.5pt}
  \fmfset{wiggly_len}{3mm}
  \fmfset{dash_len}{2.5mm}
  \fmfset{dot_size}{1.5thick}
  \fmfset{arrow_len}{2.5mm}

\begin{fmfgraph*}(80,60)
  \fmfkeep{qedneuct-3a}
  \fmfleft{gamma}
  \fmfright{in,out}
  \fmf{phantom}{out,v1,va,vertex,vb,v2,in}
  \fmf{photon}{vertex,gamma}
  \fmffreeze
  \fmf{plain}{out,v1,v2,in}
  \fmf{dashes,width=1.5pt}{v1,va,vertex,vb,v2}
  \fmfdot{vertex,v2}
  \fmfct{v1}
\fmffreeze
\fmf{plain,width=1.5pt}{v1,v2}
\end{fmfgraph*}

\begin{fmfgraph*}(80,60)
  \fmfkeep{qedneuct-3b}
  \fmfleft{gamma}
  \fmfright{in,out}
  \fmf{phantom}{out,v1,va,vertex,vb,v2,in}
  \fmf{photon}{vertex,gamma}
  \fmffreeze
  \fmf{plain}{out,v1,v2,in}
  \fmf{dashes,width=1.5pt}{v1,va,vertex,vb,v2}
  \fmfdot{v1,vertex}
  \fmfct{v2}
\fmffreeze
\fmf{plain,width=1.5pt}{v1,v2}
\end{fmfgraph*}

\begin{fmfgraph*}(80,60)
  \fmfkeep{qedneuct-6a}
  \fmfleft{gamma}
  \fmfright{in,out}
  \fmf{phantom}{out,v1,va,vertex,vb,v2,in}
  \fmf{photon}{vertex,gamma}
  \fmffreeze
  \fmf{plain}{out,v1,v2,in}
  \fmf{dashes,width=1.5pt}{v1,va,vertex,vb,v2}
  \fmfdot{v1,vertex,v2}
 \fmfrct{va}
\fmffreeze
\fmf{plain,width=1.5pt}{v1,v2}
\end{fmfgraph*}

\begin{fmfgraph*}(80,60)
  \fmfkeep{qedneuct-6b}
  \fmfleft{gamma}
  \fmfright{in,out}
  \fmf{phantom}{out,v1,va,vertex,vb,v2,in}
  \fmf{photon}{vertex,gamma}
  \fmffreeze
  \fmf{plain}{out,v1,v2,in}
  \fmf{dashes,width=1.5pt}{v1,va,vertex,vb,v2}
  \fmfdot{v1,vertex,v2}
  \fmfrct{vb}
\fmffreeze
\fmf{plain,width=1.5pt}{v1,v2}
\end{fmfgraph*}

\begin{fmfgraph*}(80,60)
  \fmfkeep{qedneuct-7}
  \fmfleft{gamma}
  \fmfright{in,out}
  \fmf{phantom}{out,v1,va,vertex,vb,v2,in}
  \fmf{photon}{vertex,gamma}
  \fmffreeze
  \fmf{plain}{out,v1,v2,in}
  \fmf{dashes,width=1.5pt}{v1,va,vertex,vb,v2}
  \fmfdot{v1,v2}
  \fmfct{vertex}
\fmffreeze
\fmf{plain,width=1.5pt}{v1,v2}
\end{fmfgraph*}

\end{fmffile}
}

\begin{figure}[t]
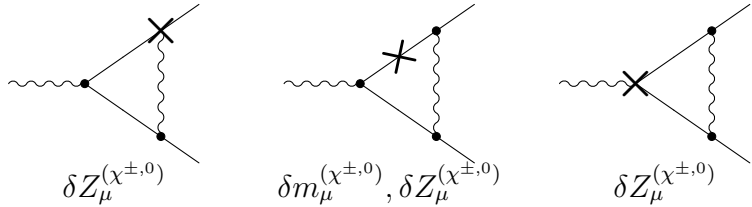

$$
\begin{array}{ccc}
  \fmfvcenter{qedct-1a} &
  \fmfvcenter{qedct-2a} &
 \fmfvcenter{qedct-5}
\\
\quad
\delta Z_{\mu} ^{(\chi^{\pm,0})}\quad
&\quad \delta m_\mu^{(\chi^{\pm,0})}, \delta Z_{\mu}^{(\chi^{\pm,0})}\quad
&\quad \delta Z_{\mu}^{(\chi^{\pm,0})}\quad
\end{array}
$$
\caption{\label{fig:ctQED} The QED counterterm diagrams and relevant
  renormalization constants. In the first and
  third diagram also $\delta e^{(\chi^{\pm,0})},
\delta Z_A^{(\chi^{\pm,0})}$ would appear, but these vanish.}
\end{figure}

Apart from the actual two-loop diagrams, two kinds of counterterm
diagrams can
contribute at ${\cal O}(\alpha n^2, \alpha c^2)$, and we denote the
corresponding contributions as $a_\mu^{\ctQED}+a_\mu^{\ctSUSY}$. The
QED counterterm diagrams  
in Fig.\ \ref{fig:ctQED}, $a_\mu^{\ctQED}$,
arise from renormalization of the QED quantities in the model
Lagrangian and involve the renormalization constants
\begin{align}
\label{QEDRenTransf}
\delta m_\mu^{(\chi^{\pm,0})},
\delta Z_\mu^{(\chi^{\pm,0})},
\delta e^{(\chi^{\pm,0})},
\delta Z_A^{(\chi^{\pm,0})}
\end{align}
for muon mass and field renormalization, charge and photon field
renormalization. 
The superscripts indicate that these QED renormalization constants
need to be evaluated at ${\cal O}(c^2,n^2)$, from SUSY one-loop
diagrams involving a chargino--sneutrino or neutralino--smuon
loop. This implies in particular 
$\delta Z_A^{(\chi^{\pm,0})}=\delta e^{(\chi^{\pm,0})}=0$. 
The values of the renormalization constants, and the physical
meaning of the renormalized parameters, are fixed by the choice of a
renormalization scheme. All QED quantities, in particular $\delta
m_\mu^{(\chi^{\pm,0})}$ and $\delta Z_\mu^{(\chi^{\pm,0})}$ must be
defined in the on-shell scheme in order to guarantee the correct
relation between the 3-point function and $a_\mu$.

\begin{figure}[t]
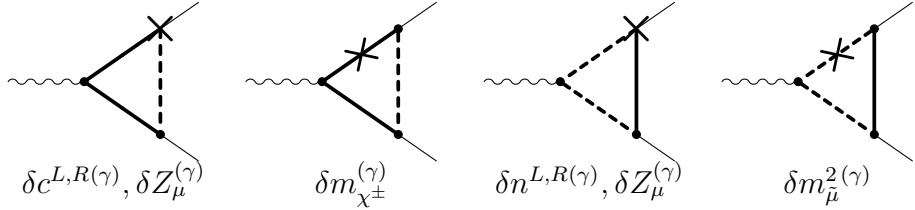

$$\begin{array}{cccc}
  \fmfvcenter{qedchact-3a}  &
 \fmfvcenter{qedchact-6a} &
  \fmfvcenter{qedneuct-3a} & 
 \fmfvcenter{qedneuct-6a}
\\
\delta c^{L,R}{}^{(\gamma)},\delta Z^{(\gamma)}_\mu
&\delta m^{(\gamma)}_{\chi^\pm}
&\delta n^{L,R}{}^{(\gamma)},\delta Z^{(\gamma)}_\mu
&\delta m_{\tilde{\mu}}^2{}^{(\gamma)}
  \end{array}
$$
\caption{\label{fig:ctSUSY} The non-vanishing SUSY counterterm
  diagrams and relevant renormalization constants.}
\end{figure}

The SUSY
counterterm diagrams in Fig.\ \ref{fig:ctSUSY}, $a_\mu^{\ctSUSY}$,
arise from renormalization of the SUSY quantities and involve the
renormalization constants 
\begin{align}
\label{SUSYRenTransf}
\delta m^{(\gamma)}_{\chi^\pm},\delta m^2_{\tilde{\mu}}{}^{(\gamma)} ,
\delta c^{L,R}{}^{(\gamma)},\delta n^{L,R}{}^{(\gamma)};\quad
\delta Z^{(\gamma)}_\mu
\end{align}
for SUSY mass and coupling renormalization and muon field
renormalization. Here all renormalization constants 
need to be evaluated at ${\cal O}(\alpha)$, i.e. from 
diagrams with a photon loop. Mass and field renormalization constants
for the neutral particles $\chi^0$, $\tilde{\nu}_\mu$ as well as
$\delta e^{(\gamma)}$ and $\delta Z_A^{(\gamma)}$ would be zero and do not
have to be included. Field renormalization
constants $\delta Z_{\chi,\tilde{\mu}}^{(\gamma)}$ could be introduced
in order to cancel UV divergences of individual two-loop
diagrams. Since they drop out in the end, we ignore them here.

Again, $\delta Z^{(\gamma)}_\mu$ has to be defined in the on-shell
scheme. With this choice the photonic (i.e.\ ${\cal O}(\alpha)$)
contribution to the muon 
field renormalization constant is UV and IR divergent, and it can be
decomposed as 
\begin{align}
\delta Z_\mu^{(\gamma)} &=
\delta Z_\mu^{(\gamma),\rm UV}+\delta Z_\mu^{(\gamma),\rm IR}
=
\frac{\alpha}{4 \pi} \left[
{- \frac{1}{\epsilon_{\text{UV}}}}
{- \frac{2}{\epsilon_{\text{IR}}}}
    - 4 + 3 {\,\rm L}(m_{\mu}^2) \right].
\end{align}
The typical transcendental constants $\gamma_E$ and $\log 4\pi$, which
cancel in the end, are ignored throughout, and ${\rm L}(m^2)$ has been
defined in Eq.\ (\ref{LDef}).

The choice of the renormalization scheme for the other parameters is
more  delicate. As discussed in Sec.\ \ref{sec:oneloop} the photonic
corrections appear as a subset of a larger class of
two-loop contributions to $a_\mu$ in models such as the
MSSM. In this context, the renormalization constants (\ref{SUSYRenTransf})
are contributions to the full renormalization constants, $\delta
m_{\chi^\pm}^{\rm full}=\delta m_{\chi^\pm}^{(\gamma)}+
\delta m_{\chi^\pm}^{\rm remainder}$ etc.
This split is not unique, and by choosing a scheme for the constants
(\ref{SUSYRenTransf}) we effectively define precisely what we mean by
``photonic corrections''. 

A natural choice for the mass counterterms are the on-shell values of
the photonic contributions to the self energies, so we choose $\delta
m^{(\gamma)}_{\chi^\pm} =
\Sigma^{(\gamma)}_{\chi^\pm}(\psl=m_{\chi^\pm})$ and
$\delta m^2_{\tilde{\mu}}{}^{(\gamma)} = 
\Sigma^{(\gamma)}_{\tilde{\mu}}(p^2=m_{\tilde{\mu}}^2)$. 
The couplings $c,n$  appear only via the combinations 
$\AA^{C,N}$ and $m_\mu\BB^{C,N}$, see Eq.\ (\ref{eq:AABBDef}). Their
renormalization transformations $c\to c+\delta c^{(\gamma)}$ and
$n\to n+\delta n^{(\gamma)}$ can be equivalently written as
\begin{align}
\AA^{C,N}&\to \AA^{C,N}+\delta \AA^{C,N}{}^{(\gamma)},\\
m_\mu\BB^{C,N}&\to m_\mu \BB^{C,N}+ m_\mu \delta \BB^{C,N}{}^{(\gamma)}
+\BB^{C,N} \delta m_\mu^{(\gamma)}.
\end{align}
In the class of models
discussed in Sec.\ \ref{sec:oneloop}, the 
combinations $\AA^{C,N}$, $\BB^{C,N}$ depend only on details of the
underlying theory, while the muon mass is a low-energy quantity. Hence
a natural choice is to renormalize 
$\AA^{C,N}$ and $\BB^{C,N}$ in the \MSbar-scheme and the muon mass in
the on-shell scheme. Implicitly, this choice defines a renormalization
scheme for the couplings $c,n$. We call the scheme defined in this way the
``on-shell muon mass scheme''.\footnote{%
In other words, in the on-shell muon mass scheme, the \MSbar-conditions
on $\AA^{C,N}$ and $\BB^{C,N}$ imply 
that $\delta(|c^L|^2+|c^R|^2)^{(\gamma)}$ is a pure \MSbar-quantity while
$\delta(c^L c^R)^{(\gamma)}$ is not (and similarly for $c\to n$). Note
that this is consistent at the 
considered order in $m_\mu$, neglecting terms of ${\cal O}(m_\mu^2)$,
but it makes difficult to individually solve for $\delta c^L$ and
$\delta c^R$ as long as it is not specified which factor among $c^L$, $c^R$ is
proportional to $m_\mu$.}

As the major advantage, in this on-shell muon mass scheme the
counterterm ${\delta(c^L c^R){}^{(\gamma)}}/({c^L c^R})$ is set equal
to ${\delta m_\mu^{(\gamma)}}/{m_\mu}+\delta\BB^C{}^{(\gamma)}/\BB^C$
and similarly for $n^L n^R$, where the first term contains a large QED 
logarithm. This is important because in the considered class of
models, $c^Lc^R$ is proportional to the muon mass, and the full 
renormalization constants will satisfy  
${\delta(c^L c^R)^{\rm full}}/({c^L c^R}) = 
{\delta m_\mu^{(\gamma)}}/{m_\mu}+\ldots$ and thus contain the same
large QED logarithm.

Rewriting the renormalization transformations in terms of differential
operators, we find a compact expression for the SUSY counterterm
contributions in our on-shell muon mass scheme,
\begin{align}
a_\mu^{\ctSUSY} &=
\delta Z_\mu^{(\gamma)}
\big(a_\mu^{\chi^\pm}+a_\mu^{\chi^0}\big)
\nonumber\\&
+\big[
{\delta m^{(\gamma)}_{\chi^\pm}}
{\partial_{
    m_{\chi^\pm}}}
+\sum_{i=L,R} \delta
c^i{}^{(\gamma,\overline{MS})}{\partial_{ c^i}} 
+\frac{\delta m_\mu^{(\gamma,\rm fin)}}{m_\mu}\BB^C{\partial_{ \BB^C}}
\big]a_\mu^{\chi^\pm}
\nonumber\\&
+
\big[
{\delta m_{\tilde{\mu}}^2{}^{(\gamma)}}
{\partial_{
    m_{\tilde{\mu}}^2}}
+\sum_{i=L,R} \delta
n^i{}^{(\gamma,\overline{MS})}{\partial_{ n^i}} 
+\frac{\delta m_\mu^{(\gamma,\rm fin)}}{m_\mu}\BB^N{\partial_{ \BB^N}}
\big]a_\mu^{\chi^0}
,
\label{amuctSUSY}
\end{align}
where it is understood that the partial derivatives $\partial_{c^i}$,
$\partial_{n^i}$ act on the couplings within $\AA^{C,N}$ and
$\BB^{C,N}$. In spite of {\em not} using the \MSbar-scheme for $c,n$
we find it convenient to express the counterterms in terms of the
would-be \MSbar-renormalization constants. For convenience, we
provide the explicit values for the renormalization constants both in
DREG and in DRED\footnote{See \cite{MartinVaughn,Mihaila} for a more general
  analysis of the transition
  from DREG to DRED.}
\begin{align}
\frac{\delta m^{(\gamma)}_{\chi^\pm}}{m_{\chi^\pm}}& = \frac{\alpha}{4\pi}
\left[-\frac{3}{\epsilon}+3{\rm L}(m_{\chi^\pm}^2)
-4-\thetaDRED
\right],\\
\frac{\delta m_{\tilde{\mu}}^2{}^{(\gamma)}}{ m_{\tilde{\mu}}^2}& = \frac{\alpha}{4\pi}
\left[-\frac{3}{\epsilon}+3{\rm L}(m_{\tilde{\mu}}^2)
-7
\right] \\
\frac{\delta n^{L,R}{}^{(\gamma,\overline{MS})}}{n^{L,R}} &=
\frac{\alpha}{4\pi}
\left[-\frac{3}{2\epsilon}
+\frac12\thetaDRED
\right] ,\\
\frac{\delta c^{L,R}{}^{(\gamma,\overline{MS})}}{c^{L,R}} &=
\frac{\alpha}{4\pi}
\left[-\frac{3}{\epsilon}
-\thetaDRED
\right] ,\\
\frac{\delta m^{(\gamma,\rm fin)}_{\mu}}{m_{\mu}}& = \frac{\alpha}{4\pi}
\left[3{\rm L}(m_{\mu}^2)
-4
\right].
\end{align}
The quantity $\thetaDRED=0$ in DREG and $\thetaDRED=1$ in the case of
DRED. 

We have checked that with these definitions all UV divergences cancel
in the sum of the two-loop diagrams of Figs.\ \ref{fig:twoloopCha}a,
\ref{fig:twoloopNeu}a and the QED counterterm diagrams with the
appropriate counterterm insertions of ${\cal
  O}(n^2,c^2)$. Likewise, all UV divergences cancel in the sum of
the two-loop diagrams of Figs.\ \ref{fig:twoloopCha}b,
\ref{fig:twoloopNeu}b and the counterterms
$a_\mu^{\ctSUSY}$. Furthermore, we have checked that the UV
divergences 
cancel for each diagram (plus corresponding counterterm diagram)
separately, if $\chi^\pm$- and $\tilde{\mu}$-field renormalization
is taken into account. 

The
IR divergences cancel only in the following combinations:
\begin{align}
  \label{eq-ergebnisse-irdivcha}
  \fmfvcenter{qedcha-5}
  + \left[ \fmfvcenter{qedct-5} \right]_{\delta Z_\mu^{(\chi^\pm)}}
  + \left[ \fmfvcenter{qedchact-3a} \right]_{\delta Z_\mu^{\rm (\gamma),IR}}
  &= {\rm fin.}\\[2ex]
  \label{eq-ergebnisse-irdivneu}
  \fmfvcenter{qedneu-5}
  + \left[ \fmfvcenter{qedct-5} \right]_{\delta Z_\mu^{(\chi^0)}}
  + \left[ \fmfvcenter{qedneuct-3a} \right]_{\delta Z_\mu^{\rm (\gamma),IR}}
  &= {\rm fin.}
\end{align}

The unsuppressed terms of order $(m_\mu/M_{\rm SUSY})^0$ cancel
between the two-loop diagrams of Figs.\ \ref{fig:twoloopCha},
\ref{fig:twoloopNeu} and the QED counterterm diagrams.

Hence, after renormalization
we obtain a UV and IR finite result of the order $(m_\mu/M_{\rm
  SUSY})^2$ for the photonic two-loop contributions to
$a_\mu$.

\section{Results}
\label{sec:results}

The full result for the photonic two-loop corrections to
$a_\mu$, defined by the sum of the two-loop diagrams of
Sec.\ \ref{sec:twoloop}, $a_\mu^{\tlgenuine}$, and the
counterterm contributions $a_\mu^{\ctQED}+a_\mu^{\ctSUSY}$, can be
cast in a quite compact 
analytical form:
\begin{align}
a^{\chipmphot}_{\mu} = \frac{1}{16\pi^2} \frac{\alpha}{4\pi}
\frac{m_\mu^2}{m_{\tilde{\nu}}^2}
\Bigg[
  & \bigg(
    \frac{1}{12}\AA^C F_1^C(x) 
    + \frac{2 m_{\chi^\pm}}{3 }  \BB^C F_2^C(x)
  \bigg) 16\log \frac{m_\mu}{m_{\tilde\nu}} 
\nonumber\\
  - &\bigg(
    \frac{47 }{72} \AA^C F_{3}^C(x) 
    + \frac{122 m_{\chi^\pm}}{9} \BB^C F_{4}^C(x)
  \bigg) 
\nonumber\\
  - & \bigg(
    \frac{1}{2} \AA^C F_1^C(x)
    + { 2m_{\chi^\pm}} \BB^C F_2^C(x)
  \bigg) {\rm L}(m_{\tilde\nu}^2) \Bigg],
\label{eq:chares}
 \\
a^{\chizphot}_{\mu} = \frac{1}{16\pi^2} \frac{\alpha}{4\pi} 
\frac{m_\mu^2}{m_{\tilde{\mu}}^2}
\Bigg[
  &\bigg(
    -\frac{1}{12} \AA^N F_1^N(x)
    +\frac{m_{\chi^0}}{3}  \BB^N F_2^N(x)
  \bigg) 16\log \frac{m_\mu}{m_{\tilde\mu}} 
\nonumber\\
  - &\bigg(
    - \frac{35 }{72}  \AA^N F_{3}^N(x)
    + \frac{16 m_{\chi^0}}{9} \BB^N F_{4}^N(x)
  \bigg) 
\nonumber\\
  + &\bigg(
    \frac{1}{4 } \AA^N F_1^N(x)
  \bigg) {\rm L}(m_{\tilde\mu}^2) \Bigg] 
\label{eq:neures}
\end{align}
where the kinematic variables are again defined as
$x=m_{\chi^\pm}^2/m_{\tilde{\nu}_\mu}^2$, or
$x=m_{\chi^0}^2/m_{\tilde{\mu}}^2$, respectively. The functions
$F_{3,4}^{C,N}$ are defined as
\begin{align}
  F_3^C(x) =\frac{4}{141 (1-x)^4} \Big[
&\big(1-x\big)\big(151 x^2-335 x + 592\big) 
\nonumber \\
   {}+ & 6 \big(21 x^3- 108x^2-93 x+50\big) \log x \nonumber\\
  - & 54 x \big(x^2-2 x-2\big) \log^2 x \nonumber\\ 
   - &108 x \big(x^2-2 x+12\big) {\rm Li}_2(1-x) \Big], \\
  F_4^C(x) = \frac{-9}{122 (1-x)^3} \Big[
&  8 \big(x^2-3 x+2\big) + \big(11 x^2 - 40x + 5\big) \log x\nonumber  \\
  -&2 \big(x^2-2 x-2\big) \log^2 x \nonumber\\
   -&4 \big(x^2-2 x+9\big) {\rm Li}_2(1-x) \Big],
\\
F_3^N(x) = \frac{4}{105 (1-x)^4} \Big[
&  \big(1-x\big)\big(-97x^2 -529x +2\big) + 6x^2\big( 13x + 81 \big) \log x \nonumber \\
  +& 108 x \big(7 x+4\big) {\rm Li}_2(1-x) \Big], \\
F_4^N(x) =\ \frac{-9}{4 (1-x)^3} \ \ \Big[
&  \big(x+3\big) \big(x \log x + x-1\big)+\big(6 x+2\big) {\rm
    Li}_2(1-x)\Big],
\end{align}
so that they are normalized to unity for $x=1$.
The logarithms in the first lines of Eqs.\ (\ref{eq:chares}),
(\ref{eq:neures})  reproduce the result of
Ref.\ \cite{DG98} for the leading QED-logarithms,
$\frac{4\alpha}{\pi}\log\frac{m_\mu}{M_{\rm SUSY}}$ times the one-loop
result. As can be easily seen,
these logarithms are negative and reduce the one-loop result by
$(7\ldots9)\%$ for SUSY masses between $100\ldots1000$~GeV. It is
interesting to note that the remaining contributions are typically
also negative and lead to a further reduction. In particular, for
$x=1$ the $F_4^C$-term alone leads to an additional $1.2\%$ reduction
of the corresponding one-loop contribution.

\begin{figure}[t]
\epsfxsize=7cm
\epsfbox{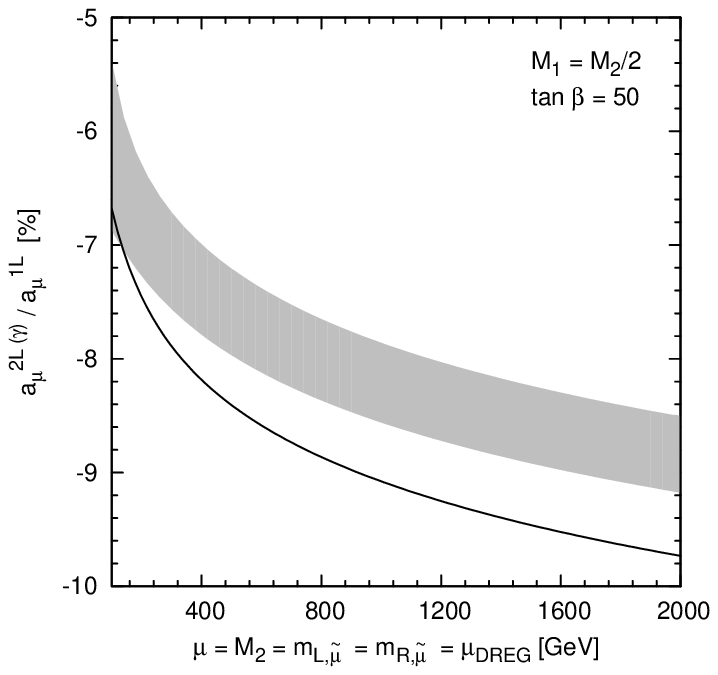}
\hfill
\epsfxsize=7cm
\epsfbox{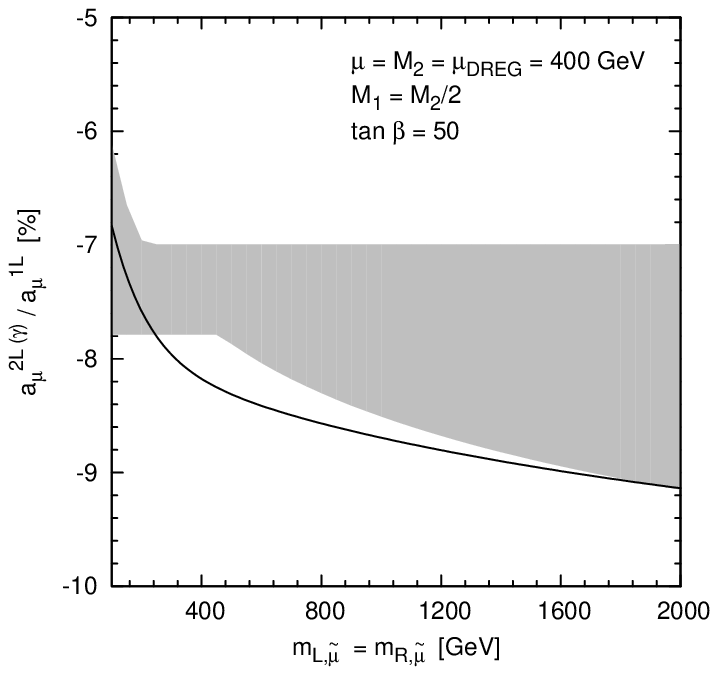}\\
\null\qquad\hfill(a)\quad\hfill\hfill\qquad(b)\ \hfill\null
\caption{\label{fig:equalmasses}
(a) Photonic two-loop corrections, relative to the MSSM one-loop
contributions, as a function of a common SUSY mass scale, see
Eq.\ (\ref{equalmasses}). 
(b) The same, as a function of $m_{L,\tilde{\mu}}=m_{R,\tilde{\mu}}$
with fixed $M_2=\mu=\mu_{\rm DREG}=400$~GeV. 
Our result is shown as a continuous line,
the leading-log approximation as a grey band.}
\end{figure}

\begin{figure}[t]
\epsfxsize=7cm
\epsfbox{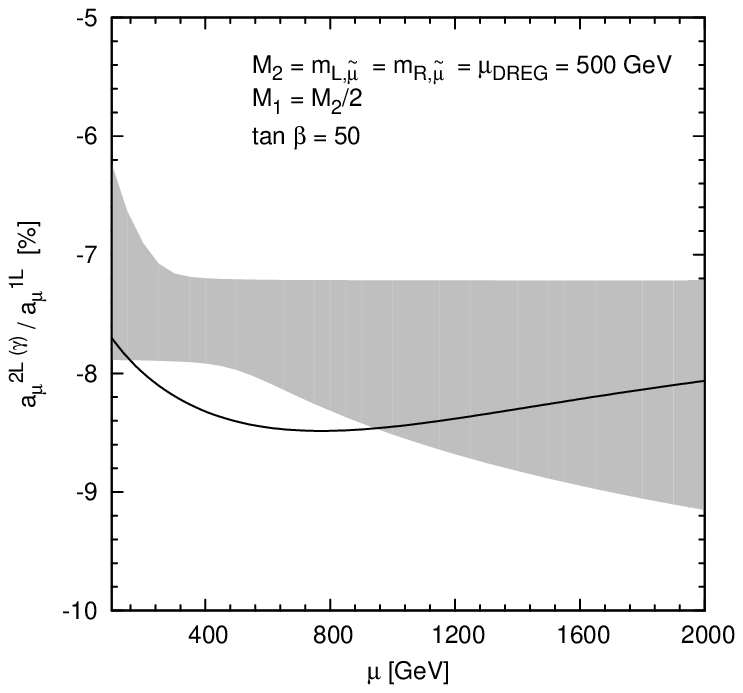}
\hfill
\epsfxsize=7cm
\epsfbox{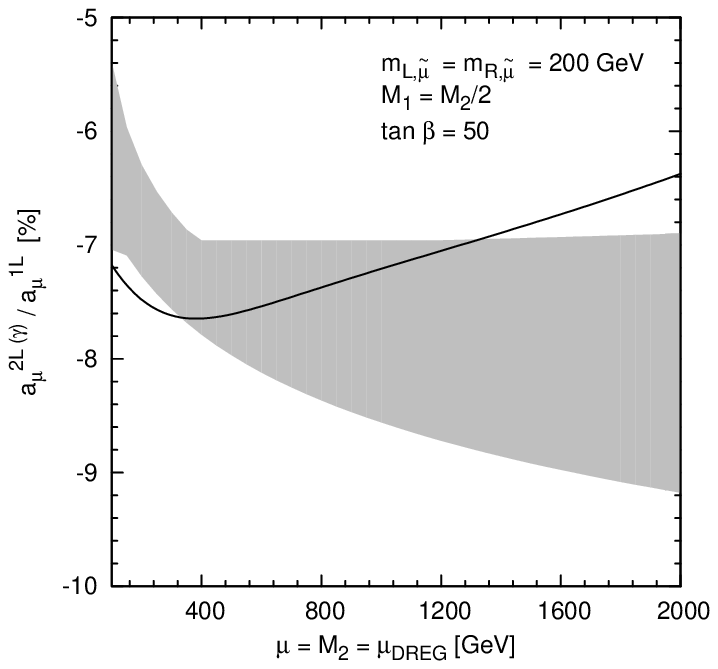}\\
\null\qquad\hfill(a)\quad\hfill\hfill\qquad(b)\ \hfill\null
\caption{\label{fig:mudep}
(a) Photonic two-loop corrections, relative to the MSSM one-loop
contributions, as a function of $\mu$. 
(b) Photonic two-loop corrections, relative to the MSSM one-loop
contributions, as a function of $\mu=M_2$. The other SUSY  parameters
are fixed as shown in the figure.}
\end{figure}

For our further discussion of the result we specialize to the case of
the MSSM. There, the one-loop and the photonic two-loop contributions
are given by
\begin{align}
\label{amuSUSY1L}
a_\mu^{\rm SUSY,1L}&=\sum_k a_\mu^{\chi^\pm}+\sum_{i,m}
a_\mu^{\chi^0},\\
\label{amuSUSYphot}
a_\mu^{\SUSYtlphot}&=\sum_k a_\mu^{\chipmphot}+\sum_{i,m}
a_\mu^{\chizphot},
\end{align}
where the SUSY masses and couplings have to be inserted
appropriately, as discussed in Sec.\ \ref{sec:oneloop}.

As a general remark, the photonic two-loop corrections depend on the
same MSSM parameters as the one-loop contributions, and owing to the
structure of the analytical results we can expect the overall
parameter dependence to be similar. In particular, the photonic
corrections are proportional to $\tan\beta$, just like the one-loop
contributions. Hence, in all our plots we choose a fixed large value
$\tan\beta=50$ and plot the ratio of the two- and one-loop
contributions.

Figure \ref{fig:equalmasses}a shows the numerical impact of the the
photonic corrections, relative to the one-loop result, for a simple,
generic case.\footnote{%
In the numerical analysis we follow
Refs.\ \cite{HSW03,HSW04,Marchetti:2008hw} and
parametrize the one-loop 
result in terms of the muon decay constant $G_\mu$, i.e.\ within
$a_\mu^{\rm SUSY,1L}$ we replace
  $\pi\alpha/s_W^2\to \sqrt2 G_\mu M_W^2$, in order to absorb further
  universal two-loop corrections. This does not influence the explicit
  factor $\alpha$ within $a_\mu^{\SUSYtlphot}$, which is
  defined in the on-shell scheme.}
We choose the fundamental SUSY
mass parameters for the Higgsino, wino, left- and right-handed smuon
equal, 
\begin{align}
\mu=M_2=m_{L,\tilde{\mu}}=m_{R,\tilde{\mu}}=\mu_{\rm DREG},
\label{equalmasses}
\end{align}
only the bino mass parameter is determined by the GUT
relation $M_1=M_2/2$, and $\tan\beta=50$. The exact result for
the ratio of the photonic corrections to the one-loop result is
denoted by the continuous line. For comparison we also show the
leading-log result, which, without further knowledge, can only be
computed with $M_{\rm SUSY}$ in a reasonable range. We choose $M_{\rm
  SUSY}$ in the range between the minimum and maximum of the mass
eigenvalues $m_{\chi^{\pm,0}}$, $m_{\tilde{\mu},\tilde{\nu}_\mu}$ and
represent the result by the grey band. 

Fig.\ \ref{fig:equalmasses}b shows a similar plot where
$M_2=\mu=\mu_{\rm DREG}=400$~GeV fixed and
$m_{L,\tilde{\mu}}=m_{R,\tilde{\mu}}$ are varied.

Not surprisingly, owing to the additional negative 
non-logarithmic contributions the exact result lies outside the band
for the leading-log estimate. The leading-log result approximates the
exact result best if we choose $M_{\rm SUSY}$ as the maximum of all
SUSY masses (because of the negative sign, this corresponds to the
lower border of the band).

In order to understand the behaviour of the MSSM result in more detail
it is important to note that the MSSM contributions are enhanced by
$\tan\beta$, but this enhancement only affects the terms involving
$\BB^{C,N}$; hence the $\AA^{C,N}$-terms are comparatively
unimportant. The $\tan\beta$-enhanced terms can be well approximated
by mass-insertion diagrams \cite{moroi,review}, with propagating
gauginos and Higgsinos and insertions of the off-diagonal entries of
the chargino/neutralino mass matrices. This implies that in the sums
over mass eigenstates in (\ref{amuSUSY1L}), (\ref{amuSUSYphot}),
intricate cancellations take place. The one-loop 
result depends sensitively on mass differences and is mainly
determined by the {\em derivatives} of the $F_i^j$ 
\cite{review}.

The same discussion can be carried out for the photonic two-loop
corrections. For the scenario with equal SUSY mass parameters the
dominant terms are the ones involving $F_2^C$ and $F_4^C$, where
$F_2^C{}'(1)=-3/4$ and $F_4^C{}'(1)=-45/122$. The smaller derivative
of $F_4^C$ partially compensates the large coefficient in
(\ref{eq:chares}), and for this reason the exact result lies only
slightly below the leading-log estimate for large $M_{\rm SUSY}$.

Fig.\ \ref{fig:mudep}a analyses the dependence on the Higgsino mass
parameter $\mu$, keeping all other SUSY mass parameters fixed. For
large or small $\mu/M_2$ there is a large spread in the chargino mass
spectrum, and the uncertainty of the leading-log estimate is large. As
the figure shows, in these parameter regions the exact result lies in
the leading-log band. 

The mass-insertion diagrams also show that the $\mu$-dependence is
non-trivial. For small $\mu$, the chargino diagrams dominate. For
large $\mu$, the diagram with bino exchange and left-right smuon
transition  can dominate \cite{review} --- it is the unique diagram
that increases linearly with $\mu$. But this behaviour is
approximately the same at the one- and two-loop level, and therefore
the ratio shown in the figure is almost constant. Nevertheless, for large
$\mu$ positive photonic contributions start to partially cancel the
leading logarithms.

Fig.\ \ref{fig:mudep}b shows the behaviour if all chargino and
neutralino masses are varied together, $\mu=M_2$ and $M_1=M_2/2$, with
fixed smuon/sneutrino mass parameters. Similarly to
Fig.\ \ref{fig:mudep}a, positive two-loop contributions 
can become important at large $\mu$, and the exact result can even lie
above the leading-log band.

\begin{figure}[t]
\begin{center}
{\includegraphics[width=.95\textwidth]{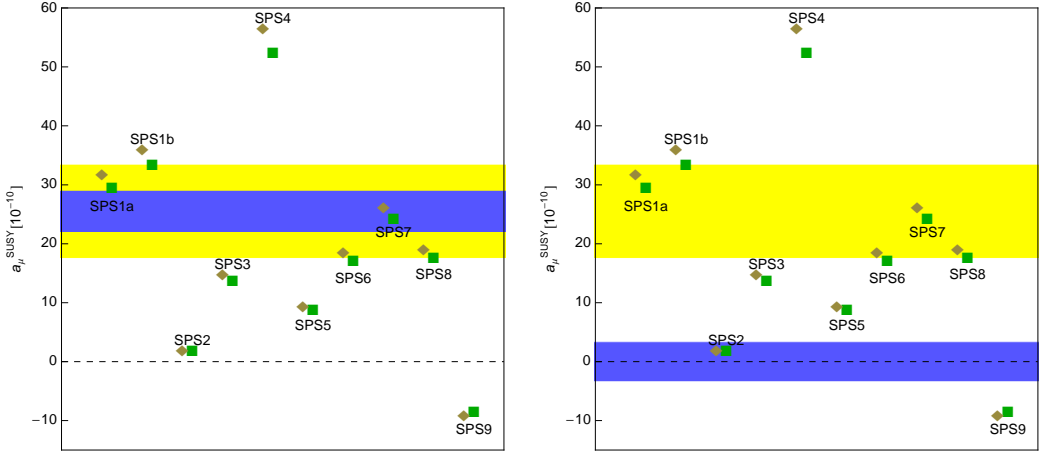}}
\caption{The predictions for $a_\mu^{\rm SUSY}$ for the Snowmass
  Points and Slopes benchmark scenarios~\cite{SPSDef}. The wide band
  corresponds to the present $1\sigma$ region of
  Eq.\ (\ref{deviation}). The narrow band 
  represents the foreseen (SM theory-limited) improved precision if
  the new $g-2$ 
  measurement is carried out \cite{FNALProposal}, given the same central
  value (left) or assuming the deviation vanishes (right). The square
  points denote the full result for $a_\mu^{\rm SUSY}$ including all
  known two-loop corrections, the diamond points are computed without
  the photonic corrections.
\label{fig:SPSplot}}
\end{center}
\end{figure}

Finally we consider the results for the SPS benchmark points
\cite{SPSDef}. E.g.\ SPS1a and SPS1b lead to $a_\mu^{\rm SUSY}$ close
to the observed deviation 
(\ref{deviation}). SPS1a has $\tan\beta=10$ and quite small SUSY
masses, SPS1b has $\tan\beta=30$ and slightly larger SUSY masses. The
results for these points are, in units of $10^{-10}$,
\begin{align*}
\begin{array}{|c|c|c|c|}
\hline
 & a_\mu^{\rm SUSY,1L} & a_\mu^{\rm SUSY,2L\ leading\ log}&
a_\mu^{\SUSYtlphot} \\
\hline
\mbox{SPS1a} & 30.49 & -1.93\ldots-2.32 & -2.18\\
\hline
\mbox{SPS1b} & 33.34 & -2.27\ldots-2.63 & -2.59\\
\hline
\end{array}
\end{align*}

Fig.\ \ref{fig:SPSplot} shows a
graphical distribution of the ten SPS benchmark predictions for
$a_\mu^{\rm SUSY}$, computed with all known one- and two-loop
corrections, and compared with the experimental and SM
value.  The discriminating power of the current and an
improved $a_\mu$ determination is evident. The plot also displays the
results obtained neglecting the photonic corrections. For several
benchmark points, the difference is as large as one sigma of the
future $\Delta a_{\mu}({\rm exp-SM})$.

\section{Matching the result to a full MSSM calculation}
\label{sec:matching}

Our results have been presented in such a way that they can be
easily utilized as building blocks in a full calculation of all two-loop
corrections to $a_\mu^{\rm SUSY,1L}$ in the MSSM. 
This is useful because two technical problems are concentrated in the
photonic corrections: the appearance of IR divergences and of terms
which are not suppressed by powers of $m_\mu$ divided by a heavy
mass of the order of the weak scale or SUSY scale.

As discussed in Secs.\ \ref{sec:introduction}, \ref{sec:oneloop}, the SUSY
two-loop corrections to SM 
one-loop diagrams are already known. What remains to be computed are
the two-loop corrections to SUSY one-loop diagrams, denoted by
$a_\mu^{\rm SUSY,2L(b)}$ in Ref.\ \cite{review}. These contributions
are defined as the two-loop diagrams where the $\mu$-lepton number is 
carried by a $\tilde{\mu}$ and/or $\tilde{\nu}_\mu$ line, plus
corresponding counterterm diagrams. 

Similar to our photonic corrections, these full contributions can be
decomposed as $a_\mu^{\tlfull} + a_\mu^{\ctQEDfull} + 
a_\mu^{\ctSUSYfull} + a_\mu^{\ctrem}$.
Here $a_\mu^{\tlfull}$ denote the genuine two-loop diagrams and
the counterterm contributions have been split into the QED counterterm 
diagrams of the form in Fig.\ \ref{fig:ctQED}, the SUSY counterterm
diagrams of the form in Fig.\ \ref{fig:ctSUSY}, and all other
remaining counterterm diagrams.  Hence we can write
\begin{align}
a_\mu^{\rm SUSY,2L(b)}&= a_\mu^{\SUSYtlphot}
\nonumber\\
&+ \big(a_\mu^{\tlfull} - \sum a_\mu^{\tlgenuine}\big)
\nonumber\\
&+ \big(a_\mu^{\ctQEDfull} - \sum a_\mu^{\ctQED}
\big)
\nonumber\\
&+ \big(a_\mu^{\ctSUSYfull} - \sum a_\mu^{\ctSUSY}\big)
\nonumber\\
&+ \big(a_\mu^{\ctrem}\big),
\label{fullsetup}
\end{align}
where the appropriate summation over the chargino, neutralino and
smuon indices of each term is implied.

Setting up the full calculation in this way has 
several technical advantages.
\begin{itemize}
\item
In the difference of the genuine loop diagrams (2nd line of
Eq.\ (\ref{fullsetup})), simply all the photonic diagrams of
Figs.\ \ref{fig:twoloopCha}, \ref{fig:twoloopNeu} drop out. In other
words, only the non-photonic two-loop diagrams need to be evaluated. 
As an advantage, these are all individually infrared finite and suppressed by
$m_\mu^2$ divided by a heavy mass squared. 
\item
The difference of the QED counterterm diagrams (3rd line of 
Eq.\ (\ref{fullsetup})) vanishes. The reason is that the appearing
QED counterterm insertions and renormalization constants, see
Eq.\ (\ref{QEDRenTransf}), have to be defined in the same way, in the
on-shell scheme, in the full theory and for the photonic
corrections. Therefore, the QED counterterm diagrams, which are again
partially infrared divergent, need not be re-evaluated in the full
calculation.  All counterterm diagrams that remain to be calculated
are individually infrared finite and suppressed by
$m_\mu^2$ divided by a heavy mass squared.
\end{itemize}

The only subtlety arises in the 4th line of Eq.\ (\ref{fullsetup}), in
connection with the SUSY counterterm diagrams of
Fig.\ \ref{fig:ctSUSY}. The renormalization constants
(\ref{SUSYRenTransf}) appearing within these diagrams are different in
the full MSSM and in 
our calculation. In the full theory, $\delta Z_\mu$ must also be
defined in the on-shell scheme, but further diagrams contribute to
it. Mass and field renormalization constants for the neutral
particles do not vanish any more. Finally, for the SUSY masses and
couplings not even the same renormalization scheme can be used in the
full theory.\footnote{%
In the MSSM, supersymmetry together with
SU(2)$\times$U(1) 
gauge invariance implies correlations between the SUSY masses and the
different couplings $c,n$, which must be reflected in the 
renormalization scheme, see e.g.\ \cite{MSSMRenorm,tf,Heidi}. But they
cannot be reflected in the purely photonic corrections since these are
not invariant under the full symmetry of the MSSM.}

Nevertheless, it is possible to choose any desired
MSSM renormalization scheme and to compute the full MSSM counterterm
diagrams $a_\mu^{\ctSUSYfull}$ in that scheme. Then simply our
counterterm result $a_\mu^{\ctSUSY}$, which has been 
given in Eq.\ (\ref{amuctSUSY}), has to be explicitly
subtracted. The choice of the regularization
scheme DREG or DRED for the counterterms has to match the choice for
the two-loop diagrams. In this way,  the full result corresponding
to the desired MSSM renormalization scheme is obtained.
As discussed in Sec.\ \ref{sec:cts}, $a_\mu^{\ctSUSYfull}$ will
contain the same large QED logarithm arising from $\delta m_\mu$ within the SUSY
coupling renormalization that is already contained in
$a_\mu^{\ctSUSY}$. Therefore, the difference in the 4th line of
Eq.\ (\ref{fullsetup}) is free of large QED logarithms.

\section{Conclusions}
\label{sec:conclusions}

In the present paper the photonic two-loop corrections to the muon
magnetic moment $a_\mu$ in the MSSM and a wider class of models have 
been evaluated exactly. The 
photonic corrections are defined as the two-loop diagrams which
contain a photon loop attached to a SUSY one-loop diagram, plus the
corresponding counterterm diagrams. The counterterms are defined in
the ``on-shell muon mass scheme'', which is natural for the models
considered here and in Ref.\ \cite{DG98} --- all QED quantities and
all masses are renormalized in the on-shell scheme, but the purely
high-scale parameters are \MSbar-renormalized.

Our result reproduces the large logarithms of Ref.\ \cite{DG98} and
provides the exact result for the additional subleading logarithms,
dilogarithms and
non-logarithmic terms. The leading logarithm has an intrinsic
uncertainty because it could be evaluated with a small, large, or intermediate SUSY
mass. For the typical SUSY scenarios considered here,
the leading logarithm amounts to around $-7\%$ of $a_\mu^{\rm
  SUSY,1L}$ if  the smallest SUSY
mass is used; the additional terms are in the range
$(0.5\ldots-2)\%$ and thus typically lead to a further reduction of
$a_\mu^{\rm SUSY,1L}$. However, for large 
$\mu$, the additional terms can have a positive 
sign and can partially compensate the leading logarithm.

It is interesting to compare the photonic two-loop corrections to
other known SUSY two-loop contributions to the muon magnetic
moment. Another universal two-loop correction is the
$(\tan\beta)^2$-correction arising from a shift of the muon Yukawa
coupling \cite{Marchetti:2008hw}. In a large part of the MSSM parameter space,
particularly for approximately degenerate SUSY masses,
the photonic and the $(\tan\beta)^2$-corrections are the largest
two-loop effects. While the photonic corrections are negative, the
$(\tan\beta)^2$-corrections are positive (for positive $a_\mu^{\rm
  SUSY,1L}$) and can overcompensate the photonic corrections for large
$\tan\beta$. The SUSY two-loop corrections to Standard Model one-loop
diagrams \cite{HSW03,HSW04} amount to around $2\%$ of $a_\mu^{\rm
  SUSY,1L}$ for degenerate masses, but in special cases with large
mass splittings or in cases where the one-loop contributions are
suppressed, they can become dominant.

Importantly, with the exact two-loop computation the theory
error arising from unknown photonic corrections has been
reduced. The remaining theory error due to unknown photonic
three-loop corrections can be estimated by comparing with the
electroweak contributions in the Standard Model. There, the photonic
three-loop corrections amount to only $1\%$ of
the photonic two-loop corrections. Based on that, we can estimate the
unknown photonic three-loop corrections to $a_\mu^{\rm SUSY,1L}$ to be
less than ${\cal O}(0.1\times10^{-10})$ and thus negligible.

Nevertheless, in order to make full use of the expected and intended
improvements of $a_\mu^{\rm exp}$ and $a_\mu^{\rm SM}$ for SUSY
phenomenology, the SUSY theory error should be further reduced. 
Our calculation can also help in computing the remaining SUSY two-loop
contributions to $a_\mu$. If the full calculation is organized as in
Eq.\ (\ref{fullsetup}), only non-photonic
diagrams and non-QED counterterm diagrams remain to be computed. These
are all infrared finite and suppressed by the required $m_\mu^2$
factor and contain no further large QED-logarithm.  

\section*{Acknowledgments} This work was supported by the German
Research Foundation DFG through Grant No.\ STO876/1-1.

\end{document}